\documentclass[nohyper,12pt,letterpaper]{JHEP3}
\usepackage{epsfig,youngtab}

\newcommand{\myfig}[3]{\begin{figure}[ht]
\begin{center}
\leavevmode \epsfxsize=#2cm \epsfbox{#1}
\end{center}
\caption{#3} \label{fig:#1}
\end{figure}}

\author{Robert de Mello Koch$^{1,2}$ Norman Ives$^{1}$ and Michael Stephanou$^{1}$\\
\qquad \\
$^{1}$ National Institute for Theoretical Physics,\\
Department of Physics and Centre for Theoretical Physics,\\ 
University of the Witwatersrand,\\ 
Wits, 2050,\\ 
South Africa\\
\qquad\\
$^{2}$Stellenbosch Institute for Advanced Studies,\\
Stellenbosch,\\
South Africa\\
\qquad\\
E-mail: \email{robert@neo.phys.wits.ac.za, Norman.Ives@students.wits.ac.za, Michael.Stephanou@students.wits.ac.za}}

\abstract{
Operators in ${\cal N}=4$ super Yang-Mills theory with an ${\cal R}$-charge of $O(N^2)$ are dual to
backgrounds which are asymtotically AdS$_5\times$S$^5$. In this article we develop efficient techniques
that allow the computation of correlation functions in these backgrounds. We find that (i) contractions 
between fields in the string words and fields in the operator creating the background are the field theory 
accounting of the new geometry, (ii) correlation functions of probes in these backgrounds 
are given by the free field theory contractions but with rescaled propagators and (iii) in these backgrounds 
there are no open string excitations with their special end point interactions; we have only closed string
excitations.
}

\preprint{WITS-CTP-039}

\title{Correlators in Nontrivial Backgrounds}

\keywords{AdS/CFT correspondence, super Yang-Mills theory}

\def \Tr{\mbox{Tr\,}}
\def \threetwo{{}^3_2}
\def \twothree{{}^2_3}

\def \twoone{{}^2_1}

\def \onethree{{}^1_3}

\def \dg{\dagger}
\def \F{ {\cal F} }

\begin{document}

\section{Introduction}

The ${1\over 2}$-BPS sector of ${\cal N}=4$ super Yang-Mills theory is a rich 
laboratory\cite{Corley:2001zk,Berenstein:2004kk,Lin:2004nb,Balasubramanian:2005mg,Brown:2006zk} for the study
of the gauge theory/gravity duality\cite{Maldacena:1997re}. This is due, in part, to the fact that
as the ${\cal R}$-charge ($J$) of an operator in the ${\cal N}=4$ super Yang-Mills theory is changed, 
its interpretation in the dual quantum gravity changes. This can be viewed as a consequence of the Myers 
effect\cite{Myers:1999ps}: as we increase $J$, the coupling to the background RR five form flux increases 
and the graviton expands. It puffs out to a radius
$$ R=\sqrt{J\over N}R_{\rm AdS},\qquad{\rm where}\qquad R_{\rm AdS}^2=\sqrt{g_{YM}^2 N} \alpha'\, . $$
We will consider the limit that $N$ is very large with $g_{YM}^2$ fixed and very small. 
For $J\sim O(1)$ the operator is dual to an object of zero size in string units, that is, a point-like 
graviton\cite{Maldacena:1997re}.
For $J\sim O(\sqrt{N})$ the operator is dual to an object of fixed size in string units - this is a string\cite{Berenstein:2002jq}.
For $J\sim O(N)$ the operator is dual to an object whose size is of the order of $R_{\rm AdS}$ - as argued in 
\cite{Balasubramanian:2001nh,Corley:2001zk} these are the giant gravitons of \cite{McGreevy:2000cw}.
The case that is of interest to us in this article is $J\sim O(N^2)$. Naively, the size of these objects diverge, even when
measured in units with $R_{\rm AdS}=1$. This divergence is simply an indication that these operators do not have an interpretation
in terms of a new object in AdS$_5\times$S$^5$: these operators correspond to new backgrounds \cite{Lin:2004nb,Balasubramanian:2005mg}.

A natural way to explore the physics of these new geometries, is to compute correlation functions in the presence of the operator
creating the new background. Since the operator creating the background has $O(N^2)$ fields, this task is non trivial. 
For the special case of operators built only from $Z$ or from $Z^\dagger$ \cite{deMelloKoch} has shown that these correlators
are easily computed using the known product rule and two point function of Schur polynomials\cite{Corley:2001zk}. These results
showed how to define operators in the super Yang-Mills theory dual to gravitons that are local in the bulk\footnote{More precisely,
they are local in the radial direction of the LLM plane and are located at $y=0$ - i.e. on the LLM plane. They are $s$-waves on both
$S^3$s in the geometry and are smeared along the $\phi$ coordinate of the LLM plane. See \cite{deMelloKoch}.} of the dual 
quantum gravity. The definition of these local operators was in terms of a modified product rule, which is a refinement of the 
usual Littlewood-Richardson rule. When using the usual Littlewood-Richardson rule, to take the product ${\tiny \yng(1)}\times R$, the 
single box would be added to all possible rows of the Young diagram $R$ as long as $R$ with the box added is again a legal Young
diagram. In contrast to this, the local operators only add boxes to a specific location in the Young diagram. Thus, for example,
we can define a local operator that would only add a box to the first row. We label these local operators by the location on the Young 
diagram to which they would add (in the case of acting with $\Tr Z$) or remove (in the case of $\Tr {d\over dZ}$) boxes. These locations
are labeled as $a_i$ (for inward point corners) and $b_i$ (for outward pointing corners) with $i$ increasing as you move along the
edge of the Young diagram from the upper right towards the lower left. See Figure 1 for an example of our labeling.
Correlators of these local operators are easily computed 
using the modified product rule\cite{deMelloKoch}. Local operators built with $O(1)$ fields, that do not mix $Z$ and $Z^\dagger$ are dual to 
gravitons; they are ${1\over 2}$ BPS probes. 

\myfig{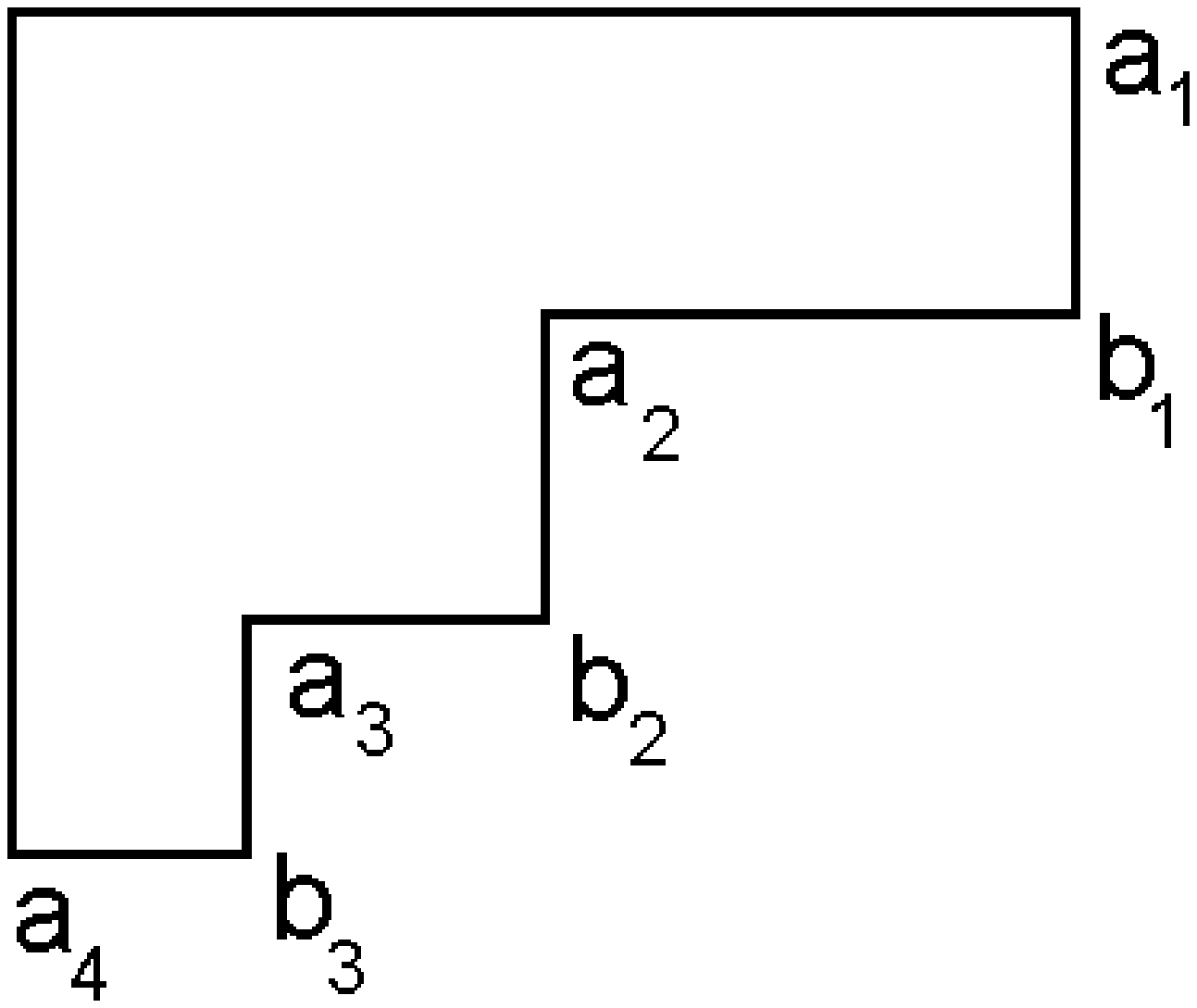}{6.5}{This figure illustrates our labeling of the corners of a Young diagram.}

Probing the background with an operator that is not ${1\over 2}$ BPS gives much richer information. In this case we have two natural
possibilities: we can excite the background by attaching an open string to obtain a restricted Schur polynomial along the lines 
of \cite{Balasubramanian:2004nb,deMelloKoch:2007uu,deMelloKoch:2007uv,Bekker:2007ea}, or we could probe the new background with closed
strings\cite{Vazquez:2006id,Chen:2007gh,deMelloKoch}. The interpretation of the open string excitation is not at all obvious. When the
${\cal R}$-charge of the operator to which the string is attached is $O(N)$, we know that the excitation indeed behaves like an open string
attached to a giant 
graviton \cite{Balasubramanian:2002sa,Balasubramanian:2004nb,Berenstein:2006qk,deMelloKoch:2007uu,deMelloKoch:2007uv,Bekker:2007ea}. These
excited giant graviton operators are the restricted Schur polynomials.
In this case the backreaction of the giant graviton can be neglected and the system is well described as a giant graviton,
with open strings attached, moving in the AdS$_5\times$S$^5$ geometry. This is nothing like the situation we study in this article. 
When the operator to which the open string is attached has an ${\cal R}$-charge of $O(N^2)$ it deforms the geometry - it is not a surface
on which open strings can end, it is a new classical geometry: a new metric with some background fluxes. Our results clearly show
that there is nothing special about how the endpoints of the string interact; they behave just like the bulk of the string. This
is a clear demonstration that there is no brane on which string endpoints end\footnote{Ofcourse, it is
possible to excite giant gravitons on these geometries, in which case open strings excitations do appear.
The perturbative string spectrum contains no open strings.}: the operator which is being excited is not a membrane;
its a new geometry. To arrive at this conclusion, we need to compute correlators of traces that mix $Z$ and $Z^\dagger$.

To probe the geometry with a closed string, one needs to compute correlators of single trace operators of the form
$$\Tr (YZ^{n_1}_{a_i} YZ^{n_2}_{a_i} Y\cdots YZ^{n_L}_{a_i}). $$
This closed string is localized at the corner $a_i$ in the geometry.
Because these operators are nearly BPS their anomalous dimensions receive only a small correction and we can safely work to one loop. 
By studying this correction, we can obtain geometric information about the new background \cite{Vazquez:2006id,Chen:2007gh,deMelloKoch}
indicating that this probe is indeed a valuable source of information about the geometry. The Wick contraction of the $Y$ fields is straight forward
because there are no $Y$s in the operator which creates the new background. After Wick contracting the $Y$ 
fields, we are left with the problem of computing correlators of traces that mix $Z$ and $Z^\dagger$. 

These mixed correlators can not be computed using the modified product rule.
In \cite{deMelloKoch} it was conjectured that these mixed correlators can be computed using modified ribbon diagrams. The modification simply
amounts to rescaling the old propagator by $c/N$, where $c$ is the weight of the box added to the background Young diagram by the
(local) operator. If true, this is a considerable simplification.

In this article we develop techniques that allow the direct computation of these correlation functions. Our results are in perfect
agreement with the conjecture of \cite{deMelloKoch}. Although we have focused on ${1\over 2}$ BPS backgrounds our results will certainly
be applicable more generally. In situations in which backreaction can be ignored, we have already developed techniques for computing the
correlation functions of restricted Schur polynomials \cite{deMelloKoch:2007uu,deMelloKoch:2007uv,Bekker:2007ea}. In these cases
contractions between fields belonging to open string words and 
the remaining fields in the restricted Schur, make a subleading contribution in
a systematic large $N$ expansion. We will argue that back reaction in the gauge theory is accounted for by including these contractions.
Our approach to computing these extra contributions starts by noting that the two point correlator (we supress spacetime dependence which 
plays no role in this article)
$$\left\langle (Z^\dagger)^k_l Z^i_j\right\rangle = \delta^i_l\delta^k_j ,$$ 
is reproduced by identifying
$$ (Z^\dagger)^k_l\leftrightarrow {d\over dZ^l_k}.$$
In this way, the contributions to a correlation function of two restricted Schur polynomials 
coming from contractions between $Z$s that belong to the open string and $Z$s that belong to the brane, can 
be written as a differential operator acting on the restricted Schur polynomials. This differential operator
will in general, contain a product of derivatives with respect to the open string words as well as derivatives
with respect to $Z$ and $Z^\dagger$. We give a rule for ``cutting'' any such product up into eight basic types 
of derivatives and then derive simple formulas for the action  of these derivatives. In this way, we can
compute arbitrary mixed trace correlators, in any background, to any order in a systematic large $N$
expansion. By specializing to the annulus geometry, we find significant simplifications allowing us to
prove the modified ribbon rule of \cite{deMelloKoch}. We then consider LLM geometries that correspond to a 
set of well seperated concentric rings. The rings give a picture of the eigenvalue density of $Z$\cite{deMelloKoch}:
the eigenvalues split into well separated clumps. In the large $N$ and large 't Hooft coupling limit 
the off diagonal modes connecting eigenvalues in different rings will be very heavy and decouple. Thus, 
$Z$ becomes block diagonal with the number of blocks matching the number of rings. Recycling the annulus
result then gives us a more general proof of the modified ribbon rule. This article is arranged as follows:
In the next section, we consider ``open string excitations'' of the annulus background. The treatment of closed 
string excitations then follows, with no extra work. In section 3 we generalize our results to backgrounds
which correspond to a set of concentric rings. In section 4 we discuss our results. The appendices collect 
some relevant background and the technical details.

\section{Backreaction: Annulus Geometry}

The calculation of two point correlation functions of restricted Schur polynomials with open strings
attached has been studied in \cite{Balasubramanian:2004nb,deMelloKoch:2007uu,deMelloKoch:2007uv,Bekker:2007ea}. In these studies,
contractions between fields in the open string and fields in the operator representing the brane were
neglected. In the present article, the number of fields in the restricted Schur polynomial is $O(N^2)$.
Operators with ${\cal R}$ charge of $O(N^2)$ are dual to new geometries, so that the back reaction of the 
operator must be taken into account. In section 2.1 we will argue that the contractions between fields in the 
open string word and the remaining fields in the operator can no longer be neglected. This is how the
backreaction of the operator on the geometry is accounted for in the gauge theory. The open string words 
that we consider will use $Z$ and $Y$ as letters. To compute correlators in the large $N$ limit, it is useful
to treat the $Y$s as defining a lattice populated by $Z$s. The $Z$s themselves can be represented by Cuntz 
oscillators, which simply keep track of the planar contractions. In this way the problem of computing
anomalous dimensions of operators becomes the problem of computing the spectrum of a Cuntz oscillator
Hamiltonian. In section 2.2 we will argue that the net effect of the backreaction is to produce a scaling of 
the Cuntz oscillators, in agreement with \cite{deMelloKoch}. A special case of this result was first obtained 
in \cite{Chen:2007gh}, for an LLM geometry with annulus boundary condition on the LLM plane. In section 2.3
we will show that the open string endpoints behave exactly like the bulk of the string. We will further
argue that the ``open string'' excitations are best thought of as closed strings propagating on a new background. 
Finally, in section 2.4 we consider probing the new backgrounds with closed strings.

\subsection{Brane/Sring Contractions}

To simplify the presentation of our methods, 
we will study an operator labeled by a rectangular Young diagram with $N_1$ rows and $M_1$
columns. Denote the irreducible representation of $S_{N_1 M_1}$ 
that this Young diagram corresponds to by $R$. We will 
consider exciting this BPS operator by attaching a single open string. The open string word has to be associated 
to the box in the\footnote{The row closest to the top is the first row; the leftmost column is the 
first column.} $N_1$th row and $M_1$th column, since this is the only box that can be removed to leave a 
valid Young diagram. Denote the irreducible representation of $S_{N_1 M_1-1}$ 
obtained by removing the box associated to the open string by $R_1$. The operator we study is
\begin{eqnarray}
\chi^{(1)}_{R,R_1}(Z,W)&=&{1\over (n-1)!}\sum_{\sigma\in S_n}\Tr_{R_1}\left(\Gamma_R(\sigma)\right) 
Z^{i_1}_{i_{\sigma (1)}}\cdots Z^{i_{n-1}}_{i_{\sigma (n-1)}}W^{i_n}_{i_{\sigma (n)}}\nonumber\\
&\equiv& \F(R,R_1)^a_b \, W^b_a .
\label{DefF}
\end{eqnarray}
For concreteness, consider an open string with a single impurity
$$ W_i^j=(Y^{n_1}ZY^{J-n_1})_i^j .$$
We assume that $J$ is $O(\sqrt{N})$ with $g_2={J^2\over N}\ll 1$ so that when contracting the open string 
words we need only sum planar diagrams\cite{Kristjansen:2002bb}. 
The correlation function we wish to compute is (attach the same open
string word to both operators)
$$ I_{R\, R_1, S\, S_1}=\left\langle \chi^{(1)}_{R,R_1}{\chi^{(1)}_{S,S_1}}^\dagger\right\rangle\, .$$
We will seperate the computation of this correlator into two pieces: $I^{(0)}_{R\, R_1, S\, S_1}$ 
obtained by neglecting contractions between the impurity in the open string word and fields in the
$\F^a_b$ piece of the operator and $I^{(1)}_{R\, R_1, S\, S_1}$ obtained by contracting the impurity 
in the open string word with a field in the $\F^a_b$ piece of the operator.

First, consider $I^{(0)}_{R\, R_1, S\, S_1}$. Using the results of \cite{deMelloKoch:2007uu}, we find
\begin{equation}
I^{(0)}_{R\, R_1, S\, S_1} = N_1 M_1 N^{J}f_R \left(1+O(g_2^2)\right)\delta_{RS}\delta_{R_1 S_1}.
\label{zero}
\end{equation}
Associate a weight $N-j+i$ to the box in the $i$th column and $j$th row of $R$. $f_R$ is the product of the weights
of the Young diagram $R$. In the language of \cite{deMelloKoch:2007uu} only the $F_0$ contraction of the open
strings contribute in the large $N$ limit of the correlator (\ref{zero}). If $R\ne S$ but $R_1=S_1$, then 
in the language of \cite{deMelloKoch:2007uu}, the only contribution comes from the $F_1$ contraction of the 
open string words. It is straight forward to consider this case using our methods, although we do not
do so in this article.

Next, consider $I^{(1)}_{R\, R_1, S\, S_1}$. After contracting all of the $Y$ fields in $W$ with the $Y^\dagger$
fields in $W^\dagger$, contract a $Z$ in $W$ with a $Z^\dagger$ in $\F^\dagger$ and a $Z^\dagger$ in $W^\dagger$
with a $Z$ in $\F$. We obtain
\begin{eqnarray} 
I^{(1)}_{R\, R_1, S\, S_1}&=& N^{J-2}\left\langle {d\F(R,R_1)^a_b\over dZ^c_d}
 {d(\F(S,S_1)^\dagger )^b_a\over d(Z^\dagger )^d_c}\right\rangle\nonumber \\
&=& N^{J-2} \left\langle {d\over dZ^c_d} {d\over d(Z^\dagger )^d_c}{d\over dW^b_a}
{d\over d(W^\dagger)_b^a}\chi^{(1)}_{R,R_1}{\chi^{(1)}_{S,S_1}}^\dagger\right\rangle \nonumber\\
&=& N^{J-2} \left\langle\Tr\left({d\over dZ}{d\over dZ^\dagger}\right)\Tr\left({d\over dW}{d\over dW^\dagger}\right)
\chi^{(1)}_{R,R_1}{\chi^{(1)}_{S,S_1}}^\dagger\right\rangle .\nonumber
\end{eqnarray}
We will now introduce a convenient graphical notation. The derivative operator that we need to consider is determined
by the fields from the open strings that are contracted with ${\cal F}$ and ${\cal F}^\dagger$. Our notation keeps
track of these fields and gives a simple picture from which we can read off the relevant derivative operator. 
We denote ${\cal F}$ and ${\cal F}^\dagger$ by open ellipses, with a single index line entering the ellipse and a single
index line leaving the ellipse. We do not draw the fields in ${\cal F}$ and ${\cal F}^\dagger$ or their contractions.
The contractions of fields in the open string words are drawn using the usual ribbon diagram (also called ``fat graph'' 
or ``double line'') representation. The $Y$ contractions are given by filled ribbons. The $Z$ contractions are empty
ribbons. Fields left uncontracted in the diagram are to be contracted with the fields in ${\cal F}$ and ${\cal F}^\dagger$.
The graphical representation of the two terms we have considered are given in figure 2.

\myfig{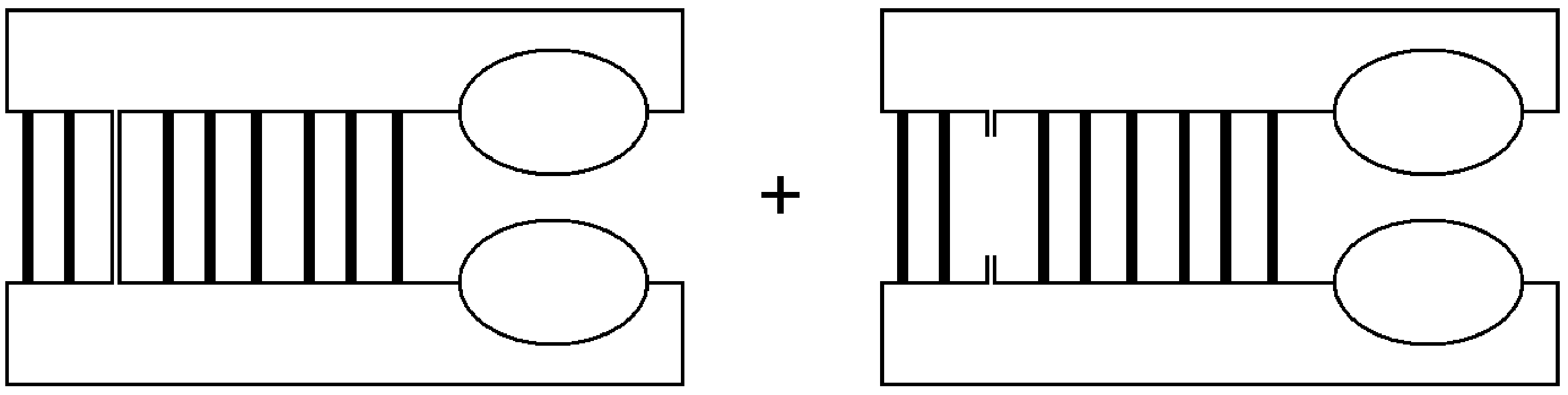}{15.0}{The graphical representation of $I^{(0)}_{R\, R_1, S\, S_1}$ and $I^{(1)}_{R\, R_1, S\, S_1}$.
We have set $n_1=2$ and $J=8$.}

To read the derivative operator from the diagram, replace each upper ``open stub'' (= uncontracted $Z$ field) by a derivative
with respect to $Z^\dagger$, each lower ``open stub'' (= uncontracted $Z^\dagger$ field) by a derivative with respect to $Z$,
the upper ellipse by a derivative with respect to the open string word $W$ and the lower ellipse by a derivative with respect 
to the open string word $W^\dagger$. All derivatives in the same index loop are in the same trace.

In general, when we have many impurities in the open string word, we may have multiple contractions between fields
belonging to the open strings and fields in $\F$ and $\F^\dagger$. In all of these cases we will be able to write these
contributions as the expectation value of a derivative operator acting on $\chi^{(1)}_{R,R_1}{\chi^{(1)}_{S,S_1}}^\dagger$.
The precise structure of the derivative operator will depend on the details of the specific contractions we consider.
As another example, if the reader translates the diagram shown in figure 3, she should obtain
$$ \left\langle\Tr\left({d^2\over dZ^2}\right)\Tr\left({d^2\over dZ^{\dagger 2}}\right)
\Tr\left({d^3\over dZ^3}{d^3\over dZ^{\dagger 3}}\right)\Tr\left({d\over dW}{d\over dW^\dagger}\right)
\chi^{(1)}_{R,R_1}{\chi^{(1)}_{S,S_1}}^\dagger\right\rangle .$$

\myfig{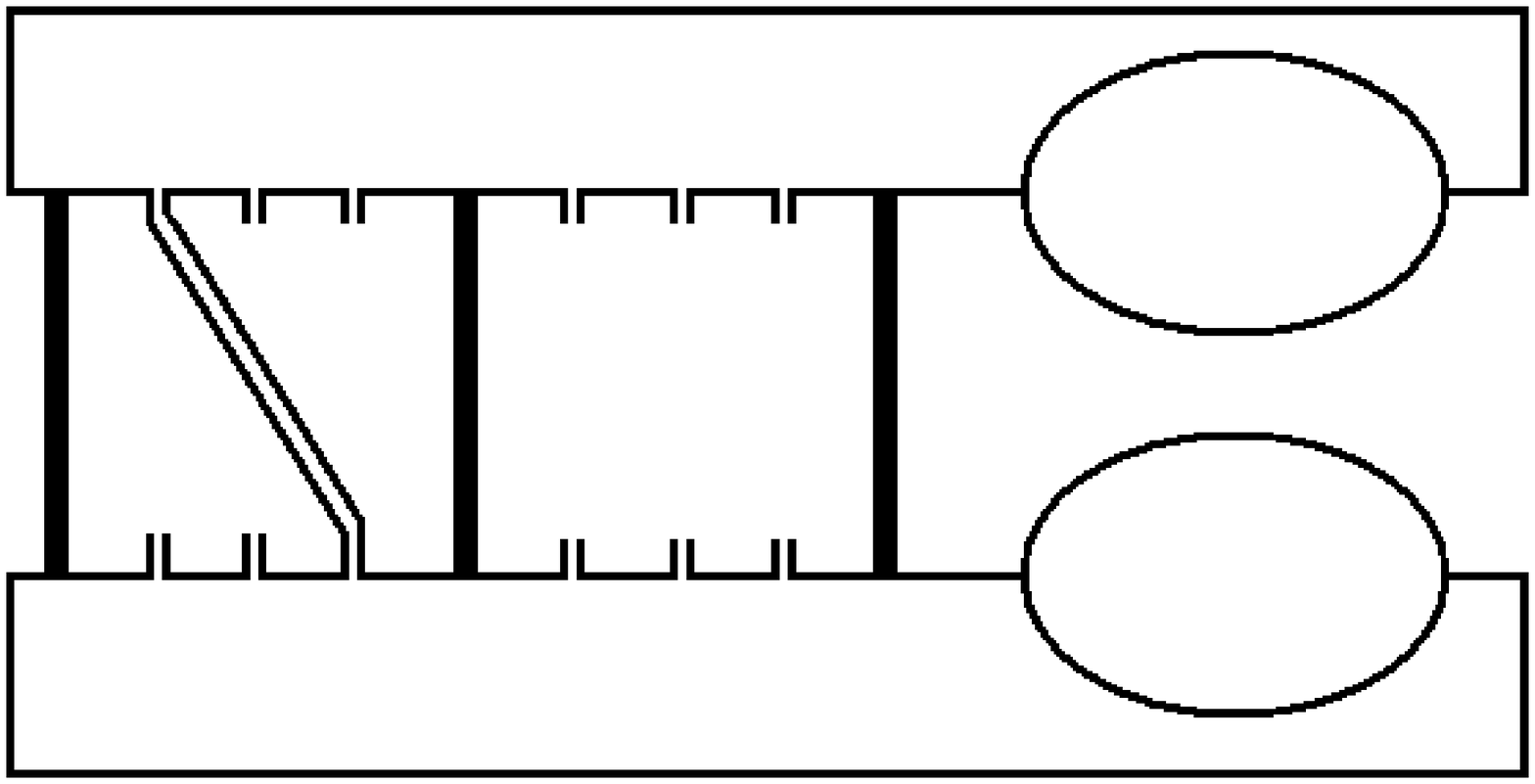}{8.0}{The graphical representation of one term contributing to the correlator 
$\left\langle \chi^{(1)}_{R,R_1}{\chi^{(1)}_{S,S_1}}^\dagger\right\rangle$.
The open string word $W=YZ^3YZ^3 Y$.}

To get the full set of contributions to the correlator we need to draw all distinct diagrams allowed such
that all possible connections of solid ribbons are included, and all possible combinations of connections
of hollow ribbons as well as disconnected stumps are included.

The fact that we can account for contractions between fields in the open string words and fields in ${\cal F}$ or
${\cal F}^\dagger$ as a derivative operator acting on the restricted Schur polynomials is a useful observation because, 
in general, we can break an arbitrary derivative operator into a product of eight basic types of derivatives, as shown 
in appendix A. We call this process ``cutting''. The first cutting rule allows us to cut single derivatives out of
any given trace\footnote{To cut a holomorphic (antiholomorphic) derivative out of the trace the derivative on its left
must also be holomorphic (antiholomorphic).} to leave a product of alternating holomorphic and antiholomorphic derivatives. 
The second rule allows us to cut the trace of a product of holomorphic and antiholomorphic derivatives
into a product of traces of purely holomorphic or purely antiholomorphic derivative. In both
cases the restricted Schur polynomial is modified by inclusion of an extra factor in the restricted character. The reader
can consult appendix A for the details. The action of these basic derivatives on a general 
restricted Schur polynomial, is described by the simple formulas collected in Appendices B and C. We call these formulas
``reduction rules''. After applying the cutting and then the reduction rules, it is straight forward to obtain
\begin{equation}
I^{(1)}_{R\, R_1, S\, S_1} = N^{J-2} (N_1 M_1)^2 f_R \left(1+O(g_2^2)\right) .
\label{one}
\end{equation}
Comparing (\ref{zero}) and (\ref{one}), we see that the contraction between the impurity in the open string and fields
in $\F(R,R_1)$ and $\F(S,S_1)^\dagger$ need only be taken into account when $N_1 M_1$ is $O(N^2 )$. This is precisely the regime
in which the operator is dual to a very heavy state whose back reaction on the original AdS$_5\times$S$^5$ space 
produces a new geometry, so it is natural to interpret these contractions as the field theory
accounting of the back reaction of the heavy state: by including these contractions, the string ``interacts with the 
back reacted geometry". This is the key result of this section, and although we have only illustrated it in a simple
example the conclusion is general.

{\vskip 0.25cm}

\noindent
{\bf Summary:} {\sl 
The contractions between fields in the open string word and the remaining fields in the operator need only be taken into 
account when the number of fields in the operator creating the background is $O(N^2 )$. 
%This operator is dual to a very
%heavy state whose back reaction produces a new geometry. 
These contractions are the field theory accounting of the back reaction of this heavy state.}

\subsection{Modified Cuntz Oscillators}

In this section, we will set $N_1=N$ and $M_1=M$. This corresponds to taking an annulus boundary condition for the dual LLM
geometry. In this case, we can have excitations of the two edges of the annulus\cite{deMelloKoch}: by acting with $Z_a$ we add 
boxes to the upper right corner of the Young diagram (corresponding to the outer edge of the annulus) and by acting with
${d\over dZ_b}$ we erode boxes from the lower right corner (corresponding to the inner edge of the annulus). 
There is a huge simplification that arises for the annulus: we can simply replace the local operators
$Z_a$ and ${d\over dZ_b}$ by $Z$ and ${d\over dZ}$. This is simply because $Z$ is unable to add boxes
anwehere except the first few rows and ${d\over dZ}$ is unable to remove boxes from anyhwere except
the last few rows. Our open string
lives at the outer edge of the annulus which implies that $R_1$ is a rectangle with $M$ columns and $N$ rows and $R$ has one extra
box in the first row, giving $M+1$ boxes in the first row.
The simplest situation in which to illustrate our result is to consider open string excitations that have multiple impurities
at a single site. For 9 impurities, the open string word is $ W^i_j=(Y(Z)^9Y)^i_j .$
We will get contributions from contracting $n$ impurities in the open string with fields in $\F,\F^\dagger$ for $n=0,1,2,...,9$.
There are $9!/(n!(9-n)!)$ distinct contractions for a given $n$. The specific details of the contractions matters. For example,
in the case that $n=4$, if none of the impurities in the open string that are contracted with $\F,\F^\dagger$ are adjacent
(see figure 4), we obtain the following contribution (this formula is correct to leading order at large $N$)
\myfig{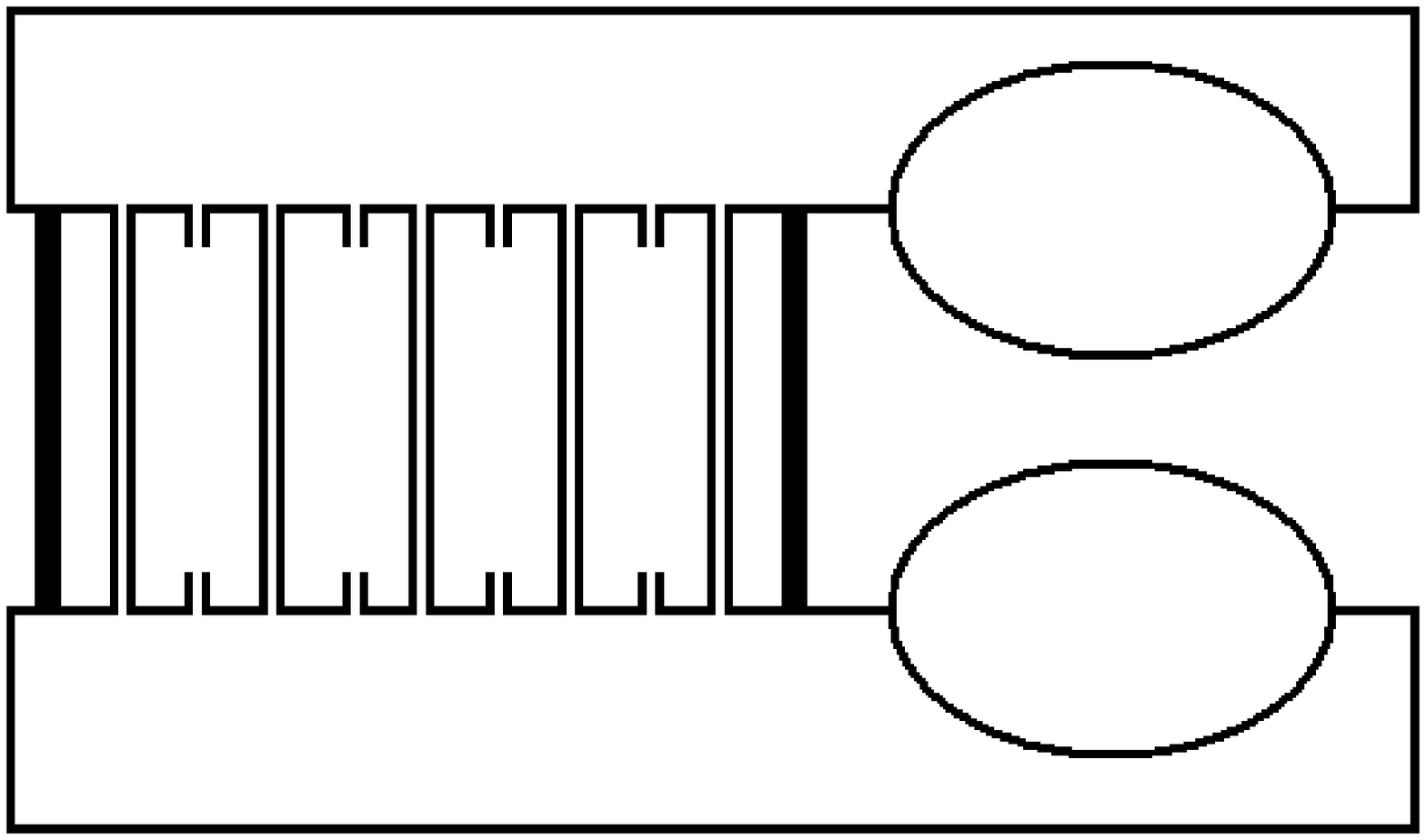}{7.5}{The diagram giving the contribution in (\ref{noadjacent})}

\begin{equation}
 N^2 \left\langle\left[\Tr\left( {d\over dZ}{d\over dZ^\dg}\right)\right]^4\Tr\left({d\over dW}{d\over dW^\dagger}\right)
\chi_{R,R_1}^{(1)}(\chi_{R,R_1}^{(1)})^\dagger\right\rangle
= (MN)^4 N^2 {M+N\over N} f_R \left(1+O(g_2^2)\right) .
\label{noadjacent}
\end{equation}

Now consider the contribution coming from the term with all four impurities adjacent (see figure 5)
\begin{eqnarray}
I_4 &=& N^5 \left\langle\Tr\left(\left[{d\over dZ}\right]^4\left[{d\over dZ^\dg}\right]^4\right)
\chi_{R,R_1}^{(1)}(\chi_{R,R_1}^{(1)})^\dagger\right\rangle\nonumber\\
&=&N^5 \left\langle\Tr\left({d\over dX_4}{d\over dX_4^\dg}\right)\prod_{i=1}^3
\Tr\left({d\over dX_i}\right)\Tr\left({d\over dX_i^\dg}\right)
\chi_{R,R_1;P}^{(1,4)}(\chi_{R,R_1;P}^{(1,4)})^\dagger\right\rangle ,
\label{alladjacent}
\end{eqnarray}
where
$$ P=(n-4,n-3,n-2,n-1).$$
To obtain this expression, we have used the methods of appendix A to decompose the derivative operator
into a product of basic types. This can now be evaluated using the methods developed in appendices B, C and D.
The details of some similar example calculations are summarized in appendix E. Although the details are completely 
different to the (\ref{noadjacent}) calculation, we find {\it exactly the same result}
\begin{equation}
I_4=N^6 M^4 {M+N\over N}f_R \left(1+O(g_2^2)\right) .
\label{finalalladjacent}
\end{equation}
\myfig{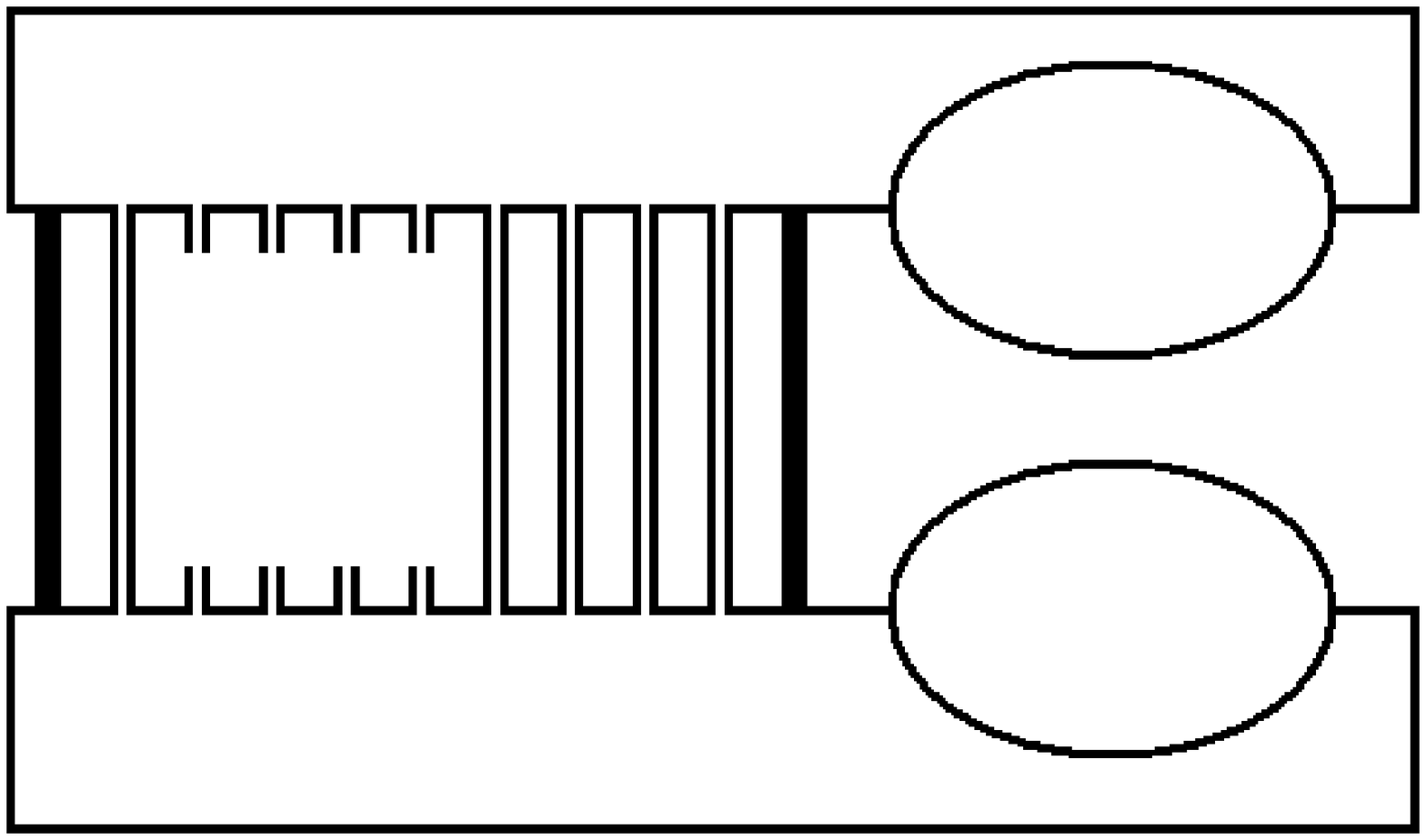}{7.5}{The diagram giving the contribution in (\ref{finalalladjacent})}

This is general: if we have $p$ impurities at a site, the contribution to the 
correlator coming from all $p!/(n!(p-n)!)$ contractions between $n$ impurities on the 
open string and the fields in $F,F^\dg$ are all the same size. Further,
it is now straight forward to check that each of the terms contributing when we have $p$ impurities
in the open string words contracting with fields in $\F$ and $\F^\dagger$, gives
$$ N^{10}\left({M\over N}\right)^p {M+N\over N}f_R $$
and therefore that
$$\langle\chi^{(1)}_{R,R_1}\chi^{(1)}_{R,R_1}{}^\dagger\rangle=\sum_{p=0}^9 N^{10}\left({M\over N}\right)^p {M+N\over N}f_R {9!\over p!(9-p)!}
=N^{9}(M+N)f_R\left( 1+{M\over N}\right)^9 .$$
If we had $n$ impurities in the site, we'd have obtained
\begin{equation}
\langle\chi^{(1)}_{R,R_1}\chi^{(1)}_{R,R_1}{}^\dagger\rangle
=(M+N)N^{n}f_R\left( 1+{M\over N}\right)^n .
\label{Nimp}
\end{equation}
Recall that adding an extra impurity in the open string word corresponds to applying another Cuntz oscillator to the state.
Clearly, in view of (\ref{Nimp}), the correct way to account for the background is to rescale the Cuntz oscillators describing 
the impurities in the open string
$$ aa^\dagger =  \left( 1+{M\over N}\right),\qquad a^\dagger a =\left( 1+{M\over N}\right)\left( 1-|0\rangle\langle 0|\right). $$
The factor $1+{M\over N}$ is ${c\over N}$ with $c$ the weight of boxes in the upper right hand region of the Young diagram.

We can give this calculation a slightly different interpretation which will allow us to state the general result: after contracting the 
$Y$ fields planarly, the above correlator can be viewed as the expectation value of a product of single trace operators, 
in the new background
\begin{eqnarray}
\nonumber
\left\langle \chi_{R,R_1}^{(1)}(Z,W)(\chi_{R,R_1}^{(1)}(Z,W))^\dagger\right\rangle
&=& \left\langle \Tr ((Z_a)^n (Z_a^\dagger)^n)\chi_{R_1}(Z)(\chi_{R_1}(Z))^\dagger\right\rangle \cr
\nonumber
&\equiv& \left\langle \Tr ((Z_a)^n (Z_a^\dagger)^n)\right\rangle_{R_1}\, .
\end{eqnarray}
These fields only add boxes in the first few rows, i.e. in the upper right region
of the Young diagram. They are thus local operators according to \cite{deMelloKoch}. 
Thus, when computing correlators in the annulus background, we can reproduce the above result by
using free field theory, after rescaling all propagators by ${c\over N}$ where $c$ is the weight of the added 
boxes. Below we will show how this generalizes for an LLM background comprised of concentric annuli.

In appendix F we give a rigorous derivation of this result. We also compute the expectation value of
$$ O=\langle \Tr \left( {d^n\over dZ^n}{d^n\over d(Z^\dagger)^n}\right)\rangle\, ,$$
for the annulus background. The result is:

{\vskip 0.5cm}

\noindent
{\bf Summary:} {\sl 
In the annulus background the original matrix $Z$ is a local operator in the sense that it only adds boxes 
in the first few rows of the Young diagram. The derivative ${d\over dZ}$ is also a local operator 
in the sense that it only removes boxes from the last few rows of the Young diagram.
To compute correlation functions of these local operators one uses 
ribbon diagrams, where each ribbon carries an extra factor of ${c\over N}$ where $c$ is the weight of the 
boxes added or removed by the local operator.}

\subsection{Tying up loose ends}

Since we are considering open string excitations, we need to pay some attention to the end point interactions. General methods to determine
the interations for a single string\cite{deMelloKoch:2007uv} or for multistrings\cite{Bekker:2007ea} are known. The strength of this interaction
is given by $\sqrt{c\over N}$. Consider a string built using $L+1$ $Y$s. These $Y$s form a lattice on which the $Z$s hop. 
The Hamiltonian for the string takes the form (this endpoint interaction assumes that the open string is 
attached to a single brane and not a boundstate of branes - see \cite{deMelloKoch:2007uv})
\begin{equation}
H = 2\lambda\sum_{l=1}^L a_l^\dagger a_l -\lambda\sum_{l=1}^{L-1}(a_l^\dagger a_{l+1}
+a_l a^\dagger_{l+1})+\sqrt{c\over N}\lambda(a_1+a_1^\dagger+a_{L}+a_{L}^\dagger).
\label{bulk}
\end{equation}
Here $c$ is the weight of the box occupied by the open string. 

If the ${\cal R}$-charge of the background is $O(1)$ or $O(\sqrt{N})$ the operator we are studying is dual to a graviton or a string, but
not a brane. In this case, the weight of the box occupied by the open string is $O(N)$, so that $\sqrt{c\over N}=1$. Further, the Cuntz
oscillators satisfy
$$ a_l a_l^\dagger =1,\qquad a_l^\dagger a_l = 1-|0\rangle\langle 0|.$$
This implies that hopping onto and off of the string is no different from hopping between bulk sites. This implies
that the end point dynamics is not special: the end points are not ``stuck to a brane''. Our string is a closed string, not an
open string. If we now consider the case of an operator with an ${\cal R}$ charge of $O(N)$ and further that the operator 
has $O(1)$ rows (or O(1) columns), then the open string is attached to a box with a weight of $\alpha N$ with $\alpha = O(1)$. 
The Cuntz oscillators are unchanged. This implies that the end point dynamics is special: hopping onto and off of the string 
has a weight $\sqrt{\alpha}$. In this case, we do indeed have an open string excitation, as has been verified 
in \cite{Balasubramanian:2002sa,Berenstein:2006qk,deMelloKoch:2007uv}. Finally, consider the case of interest
to us here, when the operator has an ${\cal R}$-charge of $O(N^2)$ and all edges with a length of $O(N)$. This 
requirement on the length of all edges is needed if the operator is to correspond to a regular LLM geometry.
\footnote{Indeed, a rectangular Young diagram with $M$ columns and $N$ rows, plus one more column with $\alpha N$ boxes
with $\alpha = O(1)$ corresponds to a finite size D3-brane on the back reacted LLM geometry. This D3 will admit open
string excitations. We are considering operators dual to geometries without any D3-branes which is achieved precisely
by our restriction that all edges have a length of $O(N)$.} 
In this case, the Cuntz oscillators are modified to 
$$ a_l a_l^\dagger ={c \over N},\qquad a_l^\dagger a_l = {c\over N}\left(1-|0\rangle\langle 0|\right)\, .$$
To make all dependence on the weight of the box occupied by the open string explicit, use the rescaled oscillators
$ a_l=\sqrt{c\over N}\tilde{a}_l.$ In terms of these oscillators
\begin{equation}
H ={c\over N}\left( 2\lambda\sum_{l=1}^L \tilde{a}_l^\dagger \tilde{a}_l -\lambda\sum_{l=1}^{L-1}(\tilde{a}_l^\dagger \tilde{a}_{l+1}
+\tilde{a}_l \tilde{a}^\dagger_{l+1})+\lambda(\tilde{a}_1+\tilde{a}_1^\dagger+\tilde{a}_{L}+\tilde{a}_{L}^\dagger)\right),
\label{bulk}
\end{equation}
$$ \tilde{a}_l \tilde{a}_l^\dagger =1,\qquad \tilde{a}_l^\dagger \tilde{a}_l = 1-|0\rangle\langle 0|\, .$$
Once again there is nothing special about the string endpoints which behave exactly like the bulk of the string!
The astute reader might object that hopping in the bulk is between two sites of the string which is different to hopping off of 
and onto the string, which is what happens at the string endpoints. This is simply an artifact of how we have split the restricted
Schur polynomial into a string plus background. Indeed as $Z$s hop off the string, extra boxes are added to the Young diagram. One could
rather describe these extra boxes as impurities in an $L+1$th site of the string. For example, if there are no extra boxes in the Young digram, 
no $Z$s can hop onto the string; with the new interpretation we would say that the $L+1$th site is empty and hence nothing can hop
out of this site. There are two facts that make this reinterpretation possible: 
\begin{itemize}
\item{} Each time we add a $Z$ in the open string word, we get an extra index loop giving an extra $N$ and an extra ${c\over N}$
        from the extra (rescaled ribbon) propagator, giving a total extra factor of $c$. By adding an extra box, the factor
        of the product of the weights ($f_R$) in the restricted Schur correlation function has an extra factor of $c$. Thus
        adding a box or an impurity contributes the same factor.  
\item{} We deal with Cuntz oscillators, that is, distinguishable particles. Thus, there are no extra $n!$ type normalizations
        that appear for $n$ bosons. Corresponding to this, the correlators of the restricted Schur polynomials is proportional
        to 1 if the Young diagrams participating have the same shape, and to 0 otherwise. (See Appendix G for a detailed matching.)
\end{itemize}
This again suggests that the excitation is best thought of as a closed string and not an open string. This has an appealing 
interpretation: the operator we are exciting has an ${\cal R}$-charge of $O(N^2)$. It does not correspond to a brane, but rather to a new 
geometry. In this case we do not expect to see any open string excitations in the spectrum. It is satisfying that this is indeed the case. 

{\vskip 0.5cm}

\noindent
{\bf Summary:} {\sl 
The dynamics of the string ``endpoints'' is identical to the dynamics of the bulk of the string. The excitation behaves like a closed string, not an
open string. This is expected since the operator being excited is dual to a new background and not a brane.}

\subsection{Back Reaction: Closed Strings}

To consider closed strings we should probe the geometry with a single trace operator
$${\cal O}=\Tr (YZ^{n_1}YZ^{n_2}Y\cdots Y Z^{n_L}). $$
The leading large $N$ contribution to this correlator is given by contracting the $Y$ fields planarly.
The above correlator then becomes the expectation value of a product of single trace operators, in the 
new background. This has been computed above.

\section{Backreaction: Multi Rings}

In this section we will consider LLM geometries that correspond to a set of well seperated thick rings. 
The background with three rings would for example, be described by a Young diagram with $N$ rows
and the same shape as the one in Figure 1. The black rings can be viewed as a picture of the eigenvalue
density of $Z$\cite{deMelloKoch}. Thus, the eigenvalues will split into three well separated clumps.
In the limit that we expect a classical geometry to emerge (large $N$ and large 't Hooft coupling) the 
off diagonal modes connecting these three subsectors will be very heavy and decouple. We expect that,
when studying almost BPS states, the effect of these modes on the dynamics can be neglected. There is
no reason to neglect off diagonal modes connecting eigenvalues in the same sector. Thus, for our
purposes, we can replace $Z$ by a block diagonal matrix with the number of blocks matching the
number of clumps of eigenvalues. If clump $i$ contains $N_i$ eigenvalues it corresponds to an
$N_i\times N_i$ block. 

A geometry with $M$ rings can thus be considered as an $M$ matrix model. The matrices $Z_i$
are $N_i\times N_i$ dimensional matrices, where clump $i$ contains $N_i$ eigenvalues. Acting
with $\Tr (Z_i)$ will only add boxes to the rows corresponding to ring $i$\cite{deMelloKoch}. These
boxes have weight $c_i$. The matrices are not interacting so that we actually have $M$ one
matrix models. Each 
of these matrix models has an annulus background - one described by a Young diagram with
$N_i$ rows. To make sure that the eigenvalues localize correctly into the multi-ring geometry,
one needs to ensure that the weight of the boxes in the rightmost column match the weights
of the corresponding boxes in the original Young diagram. This follows because the weights give
the radius squared of the position of the corresponding eigenvalue on the LLM plane\cite{deMelloKoch}.
Note that we are not just projecting the eigenvalues. Indeed, for block
$i$ we integrate over the full set of $N_i^2$ matrix elements.
We can now easily recycle the results of Appendix F to obtain
$$ {\langle\chi_B\chi_B^\dagger \Tr (Z_i^nZ_i^{\dagger\, n}) \rangle \over \langle\chi_B\chi_B^\dagger \rangle}
= N_ic_i^n\, ,$$
$$ {\langle\chi_B\chi_B^\dagger \Tr \left(
{d^n\over dZ_i^n}{d^n\over d Z_i^{\dagger\, n}}\right) \rangle \over \langle\chi_B\chi_B^\dagger \rangle}
= N_ic_i^n\, ,$$
where $c_i$ are the weights of the boxes added or removed, respectively. The computations of these correlators
is one of the main results of this article.

{\vskip 0.5cm}

\noindent
{\bf Summary:} {\sl 
In the multi-ring LLM background the original matrix $Z$ breaks into local blocks 
$Z_i$, which are $N_i\times N_i$ dimensional matrices, where clump $i$ contains $N_i$ eigenvalues.
To compute correlation functions of these local operators one uses 
ribbon diagrams, where each ribbon carries an extra factor of ${c\over N}$ where $c$ is the weight of the 
boxes added or removed by the local operator and one includes a factor of ${N_i\over N}$ for each
trace in the local operator.}

{\vskip 0.5cm}

As a nontrivial consequence of our result, note that the net affect of the background
on the Cuntz Hamiltonian (\ref{bulk}) is simply to scale the Cuntz oscillators by ${c\over N}$.

\section{Discussion}

In this article we have developed techniques which allow us to compute correlation functions in the presence of 
an operator with an ${\cal R}$-charge of $O(N^2)$. The backgrounds we have considered are LLM geometries
that correspond to a set of concentric rings. We have probed these backgrounds with operators corresponding 
to both open strings and closed strings. Contractions between fields in the string words and fields in the 
operator creating the background need only be taken into account when the number of fields in the operator 
creating the background is $O(N^2 )$; these contractions are the field theory accounting of the back reaction 
on the geometry. From the results of \cite{deMelloKoch}, we know that in the new background we can break the 
original matrix $Z$ into ``local pieces'', $Z_i$, which add boxes at specific locations on the Young diagram. 
In this 
article we have given a precise definition for this decomposition: the original $Z$ matrix decomposes into
a block diagonal matrix. There is a block for each ring. The dimension $N_i$ of the blocks is equal to the
number of eigenvalues in each ring. These blocks are the $Z_i$.
To compute correlation functions of these local operators, use the usual free field theory ribbon diagrams, but 
each ribbon now carries an extra factor of ${c\over N}$ with $c$ the weight of the boxes added by the local 
operator. The complete effect of the background is the extra ${c\over N}$ factor now carried by each 
propagator, in perfect agreement with \cite{deMelloKoch}. This is a considerable simplification.

Our study of open string excitations shows that the dynamics of the string endpoints is identical to the dynamics of the bulk of the string. 
Open string excitations of the operators with an ${\cal R}$-charge of $O(N^2)$ behave like a closed string; there are no open string excitations 
with their special end point interactions: in the new background we have only closed string excitations. This is expected since the operator being 
excited is dual to a new geometry and not a brane.

Finally, the techniques we have developed here are equally applicable to the computation of correlators in the presence of the multi-matrix 
operators of \cite{Brown:2007xh,Kimura:2007wy,Bhattacharyya:2008rb,Kimura:2008ac}.

{\vskip 0.5cm}

\noindent
{\it Acknowledgements:} We would like to thank Jeff Murugan and Joao Rodrigues for enjoyable, 
helpful discussions. This work is based upon research supported by the South African Research Chairs Initiative 
of the Department of Science and Technology and National Research Foundation. Any opinion, findings and conclusions 
or recommendations expressed in this material are those of the authors and therefore the NRF and DST do not accept 
any liability with regard thereto. This work is also supported by NRF grant number Gun 2047219.

\appendix

\section{Decomposing Derivative Operators}

As argued in section 2.1, the contributions to a correlation function of two restricted Schur polynomials, 
coming from contractions between $Z$s that belong to the open string and $Z$s that belong to the brane, can 
be written as a differential operator acting on the restricted Schur polynomials. In this appendix we will
show that any such string of derivatives can be written in terms of eight basic types of derivatives, acting
on modified restricted Schur polynomials. This result is a useful one because it is possible to work out
general formulas for the action of these eight basic derivative types on the modified restricted Schur 
polynomials. We will illustrate the basic procedure with an example, leaving a statement of the general result 
for the next section. In section A.3 we show some examples of the use of the cutting rules.

\subsection{Warm Up}

The example we study is
\begin{eqnarray}
I_2&=&\left( {d\over dZ^c_c}\right)\left({d\over dZ^d_e}{d\over dZ^e_f}{d\over d(Z^\dagger)^f_d}\right)
\left({d\over dZ^g_h}{d\over d(Z^\dagger)^h_g}\right)
\left({d\over d(Z^\dagger)^k_l}{d\over d(Z^\dagger)^l_k}\right)\times\nonumber \\
&\times&
\left({d\over dZ^a_b}{d\over dW^b_a}\right)
\left({d\over d(Z^\dagger)^m_n}{d\over d(W^\dagger)^n_m}\right)
\chi^{(1)}_{R,R_1}(Z,W)\left(\chi^{(1)}_{R,R_1}(Z,W)\right)^\dagger .
\nonumber
\end{eqnarray}
Using the notations of (\ref{DefF}), computing the derivatives with respect to the open string words gives
$$I_2=\left( {d\over dZ^c_c}\right)\left({d\over dZ^d_e}{d\over dZ^e_f}{d\over d(Z^\dagger)^f_d}\right)
\left({d\over dZ^g_h}{d\over d(Z^\dagger)^h_g}\right)
\left({d\over d(Z^\dagger)^k_l}{d\over d(Z^\dagger)^l_k}\right)
{d\F^a_b\over dZ^a_b}{d(\F^\dagger)^m_n\over d(Z^\dagger)^m_n} .$$
Computing the remaining derivatives and summing over repeated indices, we easily obtain
\begin{eqnarray}
I_2 &=&\left[{1\over (n-6)!}\right]^2\sum_{\sigma\in S_n}\sum_{\tau\in S_n}\Tr_{R_1}\left(\Gamma_R(\sigma)\right)
\Tr_{R_1}\left(\Gamma_R(\tau)\right)^* \times\nonumber\\
&\times& Z^{i_1}_{i_{\sigma(1)}}\cdots
Z^{i_{n-6}}_{i_{\sigma(n-6)}}
(Z^\dagger )^{j_1}_{j_{\tau(1)}}\cdots
(Z^\dagger )^{j_{n-6}}_{j_{\tau(n-6)}}
\delta^{i_n}_{i_{\sigma (n-1)}}
\delta^{i_{n-1}}_{i_{\sigma (n)}}
\delta^{i_{n-2}}_{j_{\tau (n-4)}}
\times\nonumber\\
&\times&
\delta^{j_{n-4}}_{ i_{\sigma (n-2)}}
\delta^{i_{n-3}}_{ i_{\sigma (n-4)}}
\delta^{j_{n-5}}_{ i_{\sigma (n-3)}}
\delta^{i_{n-4}}_{ j_{\tau (n-5)}}
\delta^{i_{n-5}}_{ i_{\sigma (n-5)}}
\delta^{j_n}_{ j_{\tau (n-1)}}
\delta^{j_{n-1}}_{ j_{\tau (n)}}
\delta^{j_{n-2}}_{ j_{\tau (n-3)}}
\delta^{j_{n-3}}_{ j_{\tau (n-2)}}\, .\label{Messy}
\end{eqnarray}
Now, define the permutations
$$ P=(n,n-1)(n-3,n-4),\qquad Q=(n,n-1)(n-2,n-3).$$
Further, set
$$ \sigma =\psi P,\qquad \tau = \lambda Q.$$
Changing variables in the above sums (\ref{Messy}) from $\sigma$ to $\psi$ and from $\tau$ to $\lambda$ we find
\begin{eqnarray}
I_2=&&\left[{1\over (n-6)!}\right]^2\sum_{\psi\in S_n}\sum_{\lambda\in S_n}\Tr_{R_1}\left(\Gamma_R(\psi P)\right)
\Tr_{R_1}\left(\Gamma_R(\lambda Q)\right)^* \times \nonumber\\
&\times& Z^{i_1}_{i_{\psi(1)}}\cdots Z^{i_{n-6}}_{i_{\psi(n-6)}}
(Z^\dagger )^{j_1}_{j_{\lambda(1)}}\cdots
(Z^\dagger )^{j_{n-6}}_{j_{\lambda(n-6)}}
\delta^{i_n}_{ i_{\psi (n)}}
\delta^{i_{n-1}}_{ i_{\psi (n-1)}}
\delta^{i_{n-3}}_{ i_{\psi (n-3)}}
\times\nonumber\\
&\times&
\delta^{i_{n-5}}_{ i_{\psi (n-5)}}
\delta^{j_n}_{ j_{\lambda (n)}}
\delta^{j_{n-1}}_{ j_{\lambda (n-1)}} 
\delta^{j_{n-2}}_{ j_{\lambda (n-2)}}
\delta^{j_{n-3}}_{ j_{\lambda (n-3)}}
\delta^{i_{n-2}}_{ j_{\lambda (n-4)}}
\delta^{j_{n-4}}_{ i_{\psi (n-2)}}
\delta^{j_{n-5}}_{ i_{\psi (n-4)}}
\delta^{i_{n-4}}_{ j_{\lambda (n-5)}}\, .\nonumber
\end{eqnarray}
The reason why we made the change of variables from $\sigma$ and $\tau$ to $\lambda$ and $\psi$ is now clear: 
in (\ref{Messy}) Kronecker deltas with two $i$ indices or two $j$ indices did not have the property that the upper
index was related to the lower index by permutation; after the change of variables, all such Kronecker deltas
do have this property. This is useful, because a Kronecker delta with this property is produced by acting on the 
restricted Schur polynomial with the trace of a derivative. One is tempted to replace all such Kronecker deltas
with indices $i_j$ $j<n$ by the trace of a derivative with respect to $Z$; this is not quite correct. As an example,
$\delta^{i_{n-1}}_{ i_{\psi (n-1)}}$ in the last expression above is obtained by differentiating only 
$Z^{i_{n-1}}_{i_{\psi (n-1)}}$ - the trace of a derivative with respect to $Z$ will generate this term as well
as terms that come from acting on every single other $Z$ in the polynomial. Further, due to the prescence of $P$ and
$Q$ it really does make a difference which $Z$ is differentiated. This is, however, easily overcome: we
can replace $Z^{i_{n-1}}_{i_{\psi (n-1)}}$ by a new matrix $X^{i_{n-1}}_{i_{\psi (n-1)}}$ so that 
$\delta^{i_{n-1}}_{ i_{\psi (n-1)}}$ can safely be replaced by the trace of a derivative with respect to $X$.
We call these new matrices ``open string place holders''. 
It is easy to see that $I_2$ now takes the form
$$ I_2=
\Tr {d\over dX_1}
\Tr {d\over dX_3}
\Tr {d\over dX_5}
\Tr {d\over dW}
\Tr {d\over dX_1^\dagger}
\Tr {d\over dX_2^\dagger}
\Tr {d\over dX_3^\dagger}
\Tr {d\over dW^\dagger}
\Tr {d\over dX_2}{d\over dX_4^\dagger}\times$$
$$\times \Tr {d\over dX_4}{d\over dX_5^\dagger}
\chi^{(1,5)}_{R,R_1;P}(Z,W)(\chi^{(1,5)}_{R,R_1;Q}(Z,W))^\dagger, $$
where we have introduced the new notation
$$\chi^{(1,m)}_{R,R_1;\Lambda}(Z,W)\equiv
{1\over (n-1)!}\sum_{\sigma\in S_n}\Tr_{R_1}\left(\Gamma_R(\sigma \Lambda)\right) 
Z^{i_1}_{i_{\sigma (1)}}\cdots Z^{i_{n-m-1}}_{i_{\sigma (n-m-1)}}
\prod_{k=1}^m X^{i_{n-k}}_{i_{\sigma (n-k)}}W^{i_n}_{i_{\sigma (n)}}\, .$$
In this formula $\Lambda$ is any element of the symmetric group.
Thus, the original derivative operator has been decomposed into a product of basic operations
as advertised. The Schur polynomial has been modified by the inclusion of a new factor ($\Lambda$
in the last equation) inside the trace; we call this factor the {\sl trace insertion}.
Since the trace insertion is a new factor in the trace, our notation includes the trace insertion 
after the existing trace labels.

\subsection{General Rule}

In this section we give general rules for decomposing a differential operator into a product
of basic operations. The full set of basic operations is
$$\Tr \left( {d\over dZ}\right),\quad \Tr\left( {d\over dW}\right),\quad \Tr\left( {d\over dZ^\dagger}\right),
\quad \Tr\left( {d\over dW^\dagger}\right),\quad 
\Tr \left( {d\over dZ}{d\over dZ^\dagger}\right),\quad $$
$$ \Tr \left( {d\over dW}{d\over dZ^\dagger}\right),\quad
\Tr \left( {d\over dZ}{d\over dW^\dagger}\right)\quad {\rm and}\quad \Tr \left( {d\over dW}{d\over dW^\dagger}\right).$$
We call the last four operators ``mixed derivatives". 

A general rule must give a recipe for reading off the trace insertion and product of basic operations (the new derivative
operator) from any differential operator to be disected. Of course, it is just a summary of what happens when one performs the analog
of the $\sigma,\tau$ $\to$ $\psi,\lambda$ change of variables of the last section.  

In this section, we assume that the open string word is associated with the n$th$ index $i_n$ as in (\ref{DefF}). In what follows
we will switch to an obvious matrix notation, illustrated in the following example
$$ {d\over dZ^a_b} {d\over dW^b_c}{d\over d(Z^\dagger )^c_d}{d\over d(W^\dagger )^d_a}\to (D D_W D^\dg D_W^\dg )\, . $$
Terms within a single bracket are traced. We start by giving each of the derivatives with respect to $W$ or $Z$ a label, 
counting down from $n$. $D_W$ is given the label $n$. We then give each of the derivatives with respect to $W^\dagger$ or 
$Z^\dagger$ a label, again counting down from $n$. $D_W^\dagger$ is given the label $n$. As an example, the operator
$$(D)(DDD^\dg )(D D^\dg )(D^\dg D^\dg )(D D_W )(D^\dg D_W^\dg )$$
is labelled as follows (the labels for $D,D_W$ appear above the operator; the labels for $D^\dg,D_W^\dg$ appear 
below the operator) 
$$ \matrix{n-5 \cr (D)\cr .}\,\, \matrix{n-4 &n-3 &. \cr (D &D &D^\dg)\cr . &. &n-5}
\,\, \matrix{n-2 &.\cr (D &D^\dg)\cr . &n-4}\,\, \matrix{. &.\cr (D^\dg &D^\dg)\cr n-3 &n-2}
\,\,\matrix{n-1 &n\cr (D &D_W)\cr . &.}\,\,\matrix{. &.\cr (D^\dg &D_W^\dg)\cr n-1 &n}.$$
The $Z$ derivatives with labels will be replaced with open string place holders. There are two cutting rules:

{\vskip 0.5cm}

{\sl First cutting rule:} If, within any given trace, $D$ (or any other holomorphic derivative)
has another holomorphic derivative to its left, it can
be removed from the trace and placed into its own trace. The two cycle which swaps the label of $D$ and the label
of its neighbour on the left is added, on the left, to the trace insertion of the holomorphic Schur polynomial.
If, within any given trace, $D^\dg$ (or any other antiholomorphic derivative)
has another antiholomorphic derivative to its left, it can
be removed from the trace and placed into its own trace. The two cycle which swaps the label of $D^\dg$ and the label
of its neighbour on the left is added, on the left, to the trace insertion of the antiholomorphic Schur polynomial.

{\vskip 0.25cm}

{\sl Second cutting rule:} If within any given trace $DD^\dg$ (or any other product of a holomorphic with an
antiholomorphic derivative) has a second $DD^\dg$ (or any other product of a holomorphic with an
antiholomorphic derivative) to its right, then the ``middle two'' derivatives can be removed from the existing
trace and placed into their own trace. The two cycle which swaps the labels of the two holomorphic derivatives
is added, on the left, to the trace insertion of the holomorphic Schur polynomial.
If within any given trace $D^\dg D$ (or any other product of an antiholomorphic with a
holomorphic derivative) has a second $D^\dg D$ (or any other product of an antiholomorphic with a
holomorphic derivative) to its right, then the ``middle two'' derivatives can be removed from the existing
trace and placed into their own trace. The two cycle which swaps the labels of the two antiholomorphic derivatives
is added, on the left, to the trace insertion of the antiholomorphic Schur polynomial.

{\vskip 0.5cm}

We have stated the rules using the terms ``holomorphic/antiholomorphic'' derivative. Stated in this way, the rule
are valid even if there is more than one open string attached to the restricted Schur polynomial. Any derivatives 
cut out of the product, with respect to $Z$ or $Z^\dagger$ are replaced by derivatives with respect to open string 
place holders. 

\subsection{Examples}

In this appendix we give some examples of how the cutting rules are used. This is done so that the reader can test
that she understands how to correctly apply the rules. The operator
$$\Tr\left({d\over dZ}{d\over dZ}{d\over dZ}{d\over dW}\right)$$
becomes 
$$\Tr\left({d\over dX_3}\right)\Tr\left({d\over dX_2}\right)\Tr\left({d\over dX_1}\right)\Tr\left({d\over dW}\right).$$
The antiholomorphic trace insertion is $1$; the holomorphic trace insertion is $(n-3,n-2,n-1,n)$. The operator
$$\Tr\left({d\over dZ}{d\over dZ^\dg}{d\over dZ}{d\over dZ^\dg}{d\over dZ}{d\over dZ^\dg}\right)$$
becomes 
$$\Tr\left({d\over dX_3}{d\over dX_2^\dg}\right)\Tr\left({d\over dX_2}{d\over dX_1^\dg}\right)\Tr\left({d\over dX_1}{d\over dX_3^\dg}\right).$$
The antiholomorphic trace insertion is $1$; the holomorphic trace insertion is $(n-3,n-2)(n-1,n-2)$. By cycling a derivative around the
operator we have dissected can be written as
$$\Tr\left({d\over dZ^\dg}{d\over dZ}{d\over dZ^\dg}{d\over dZ}{d\over dZ^\dg}{d\over dZ}\right)$$
Cutting this operator up gives a holomorphic trace insertion of $1$ and a nontrivial antiholomorphic trace insertion. Clearly the result
of cutting is not unique. Of course, these different dissections all lead to the same value for the correlation function.

\section{Mixed Derivative Rules}

In this appendix we will explain how to evaluate
$$\left\langle \left[\hat{O}\chi^{(1)}_{R,R_1}(Z,W)(\chi^{(1)}_{S,S_1}(Z,W'))^\dagger\right]\right\rangle $$
in free field theory, in the case that $\hat{O}$ is one of the mixed derivative operators. All the arguments in this appendix are unchanged
if a trace insertion factor is included.

\subsection{$\hat{O}=\Tr \left( {d\over dZ}{d\over dZ^\dagger}\right)$}

Consider the Schwinger-Dyson equation
$$0=\int dZdZ^\dagger dYdY^\dagger {d\over dZ^a_b}\left(
\chi^{(1)}_{R,R_1}(Z,W)\left[{d\over d(Z^\dagger)^b_a}(\chi^{(1)}_{S,S_1}(Z,W'))^\dagger\right] e^{-S}\right),$$
where
$$ S=\Tr (ZZ^\dagger +YY^\dagger).$$
The Schwinger-Dyson equation implies
\begin{eqnarray}
\left\langle \left[\Tr \left( {d\over dZ}{d\over dZ^\dagger}\right)
\chi^{(1)}_{R,R_1}(Z,W)(\chi^{(1)}_{S,S_1}(Z,W'))^\dagger\right]\right\rangle &=&
\left\langle \chi^{(1)}_{R,R_1}(Z,W)(Z^\dagger)^b_a
\left[{d\over d(Z^\dagger)^b_a}(\chi^{(1)}_{S,S_1}(Z,W'))^\dagger\right]\right\rangle
\nonumber\\
&=& n_{Z^\dagger}\left\langle
\chi^{(1)}_{R,R_1}(Z,W) (\chi^{(1)}_{S,S_1}(Z,W'))^\dagger\right\rangle
\nonumber
\end{eqnarray}
where $n_{Z^\dagger}$ is the number of $Z^\dagger$ matrices appearing in $(\chi^{(1)}_{S,S_1}(Z,W'))^\dagger $.
The correlator $\langle\chi^{(1)}_{R,R_1}(Z,W) (\chi^{(1)}_{S,S_1}(Z,W'))^\dagger\rangle$ is now easily evaluated
using the results of \cite{deMelloKoch:2007uu}.

\subsection{$\hat{O}=\Tr \left( {d\over dW}{d\over dW^\dagger}\right)$}

This operator simply ``contracts" the two open string words - it picks out the $F_0$ contribution to the correlator
in the language of \cite{deMelloKoch:2007uu}. Thus,
$$\left\langle \Tr \left( {d\over dW'}{d\over dW^\dagger}\right)
\chi^{(1)}_{R,R_1}(Z,W')(\chi^{(1)}_{S,S_1}(Z,W))^\dagger\right\rangle $$
is simply equal to the coefficient of the $F_0$ contribution to the correlator 
$$\left\langle\chi^{(1)}_{R,R_1}(Z,W') (\chi^{(1)}_{S,S_1}(Z,W))^\dagger\right\rangle\, .$$

\subsection{$\hat{O}=\Tr \left( {d\over dZ}{d\over dW^\dagger}\right)$}

Explicitely performing the derivative with respect to $Z$ in
$$I={d\over dZ^e_d}\chi^{(1)}_{R,R_1}(Z,W){d\over d(W^\dagger)_e^d}(\chi^{(1)}(Z,W))^\dagger $$
we obtain
\begin{eqnarray}
I &=&{1\over (n-2)!}\sum_{\sigma\in S_n}\Tr_{R_1}\left(\Gamma_R(\sigma )\right) Z^{i_1}_{i_{\sigma (1)}}
\cdots Z^{i_{n-2}}_{i_{\sigma (n-2)}}\delta_e^{i_{n-1}}\delta^d_{i_{\sigma (n-1)}}W^{i_n}_{i_{\sigma (n)}}
{d\over d(W^\dagger )^d_e}(\chi^{(1)}(Z,W))^\dagger\nonumber\\
&=&{1\over (n-2)!}{d\over dX^e_d}\sum_{\sigma\in S_n}\Tr_{R_1}\left(\Gamma_R(\sigma )\right) Z^{i_1}_{i_{\sigma (1)}}
\cdots Z^{i_{n-2}}_{i_{\sigma (n-2)}}X^{i_{n-1}}_{i_{\sigma (n-1)}}W^{i_n}_{i_{\sigma (n)}}
{d\over d(W^\dagger )^d_e}(\chi^{(1)}(Z,W))^\dagger\, .
\nonumber
\end{eqnarray}
If we now introduce the representations $T_\alpha$ defined by removing a single box from $R_1$, so that
$$R_1=\oplus_\alpha T_\alpha ,$$
we obtain
$$ I=\sum_\alpha {d\over dX^e_d}\chi^{(2)}_{R,T_\alpha}(Z,X,W){d\over d(W^\dagger)_e^d}(\chi^{(1)}(Z,W))^\dagger ,$$
where in the restricted Schur polynomial $\tilde{\chi}^{(2)}_{R,T_\alpha}(Z,X,W)$, $W$ is associated with the box that must be
removed from $R$ to obtain $R_1$ and $X$ is associated with the box that must be removed from $R_1$ to obtain $T_\alpha$. After
using the subgroup swap rule of \cite{deMelloKoch:2007uu} 
to swap $X$ and $W$, this correlator can be evaluated exactly as in the previous subsection.

\subsection{$\hat{O}=\Tr \left( {d\over dW}{d\over dZ^\dagger}\right)$}

The evaluation of this term is essentially the same as the term treated in the last subsection.

\section{Reduction Rules}

In this section we will consider the action of
$$ \Tr\left({d\over dZ}\right)\equiv D_Z,\quad {\rm and} \quad \Tr\left({d\over dW}\right)\equiv D_W,$$
on restricted Schur polynomials. By $D_W$ we mean either a reduction with respect to the open string attached to 
the restricted Schur polynomial or with respect to any of the open string place holders.
We call these ``reductions'' of the restricted Schur polynomial because the action
of the operators removes boxes from the Young diagram label of the polynomial. The action of $D_W$ on a 
restricted Schur polynomial has been worked out in \cite{deMelloKoch:2007uu}. $D_W$ removes the box associated
with $W$, thereby producing a Schur polynomial and multiplies this polynomial by the weight of the removed
box. 

Now, consider the action of $D_Z$. If $D_Z$ acts after $D_W$ has acted, we need the action of $D_Z$ on a Schur 
polynomial. This action has been worked out in \cite{deMelloKoch:2004ws} and \cite{deMelloKoch:2007uu}. $D_Z$
when acting on a Schur polynomial produces all Schur polynomials that can be obtained by removing a single
box from the Schur polynomial it acts on. Each of the polynomials produced are multiplied by the weight
of the removed box. 

Finally, we will evaluate the action of $D_Z$ on a restricted Schur polynomial. By explicitely evaluating the
derivative, we have
\begin{eqnarray}
{d\over dZ^a_a}\chi^{(1)}_{R,R_1}(Z,W)&=&
{1\over (n-2)!}\sum_{\sigma\in S_n}\Tr \left(\Gamma_R(\sigma)\right) Z^{i_1}_{i_{\sigma (1)}}\cdots
Z^{i_{n-2}}_{i_{\sigma (n-2)}}\delta^{i_{n-1}}_{i_{\sigma (n-1)}}W^{i_n}_{i_{\sigma (n)}}\nonumber\\
&=&D_X\sum_{\alpha}\chi^{(2)}_{R,T_\alpha}(Z,X,W) ,\label{Xred}
\end{eqnarray}
where in the restricted Schur polynomial $\chi^{(2)}_{R,T_\alpha}(Z,X,W)$, $W$ is associated with the box that must be
removed from $R$ to obtain $R_1$ and $X$ is associated with the box that must be removed from $R_1$ to obtain $T_\alpha$.
In this last formula, the representations $T_\alpha$ are all representations that can be obtained by removing a single box from 
$R_1$, so that
$$ R_1=\oplus_\alpha T_\alpha . $$
The reduction with respect to $X$ in (\ref{Xred}) is now easily computed using the subgroup swap rule of \cite{deMelloKoch:2007uu}.
Clearly, the arguments in this appendix are unchanged if a trace insertion factor is included.

\subsection{Example}

For this subsection we will use a graphical notation for the labels of the restricted Schur polynomial. We draw $R$ as a Young diagram
and write the open string word $w$ in the box which must be removed to obtain $R_1$. Similarly, we write $x$ into the box that must
be removed to obtain $T_\alpha$. In this notation, an explicit example of (\ref{Xred}) is 
$$ D_Z \chi_{\small \young({\,}{\,}{w},{\,})}=D_x\left(
\chi_{\small \young({\,}{\,}{w},{x})}+\chi_{\small \young({\,}{x}{w},{\,})}
\right).$$
We can simply evaluate the action of $D_x$ because when the polynomial is constructed we first reduce with to the $w$ box and then
with respect to the $x$ box; we need to swap these two using the subgroup swap rule.
To apply the subgroup swap rule, we do not need to worry about twisted string states because we are reducing with respect to $x$
(see \cite{deMelloKoch:2007uu}). Thus, after swapping we obtain
$$D_x\left(
{1\over 9}\chi_{\small \young({\,}{\,}{x},{w})}+{8\over 9}\chi_{\small \young({\,}{\,}{w},{x})}
+\chi_{\small \young({\,}{w}{x},{\,})}
\right).$$
To reduce with respect to $x$ we now simply remove the box populated by $x$ and multiply by its weight so that we finally obtain
$$ D_Z \chi_{\small \young({\,}{\,}{w},{\,})}=
{N+2\over 9}\chi_{\small \young({\,}{\,},{w})}+{8(N-1)\over 9}\chi_{\small \young({\,}{\,}{w})}
+(N+2)\chi_{\small \young({\,}{w},{\,})}.$$

\section{Formulas for Restricted Characters}

The cutting rules introduce an insertion factor for each restricted Schur polynomial in the correlator.
Evaluating this extra factor is most easily done using restricted characters. In \cite{Bekker:2007ea} 
general formulas for restricted characters were obtained. In this appendix we will review these methods.
In the next appendix we illustrate our methods with a nontrivial example.

A restricted character is given by taking a restricted trace of a group element. By a restricted trace,
we mean that we don't trace over the whole carrier space on which the group acts; we trace only over a subspace
$$ \chi_{R,R_1}(\sigma)=\Tr_{R_1}\left(\Gamma_R(\sigma)\right).$$
$R$ is an irrep of $S_n$; we can think of $R$ as a Young diagram with $n$ boxes.  The subspace $R_1$ is the 
carrier space of a subgroup of $S_n$. Consequently, a convenient way to specify which subspace of the full 
space we consider, is by knocking boxes off the Young diagram $R$; the smaller Young diagram is $R_1$. Finally,
we also need to consider restricted characters in which the row and column indices are traced over different 
subspaces. In this case, we compute
$$ \chi_{R,R_1 R_2}(\sigma)=\Tr_{R_1 R_2}\left(\Gamma_R(\sigma)\right)$$
by summing the row index over $R_1$ and the column index over $R_2$. This requires that we have an isomorphism
between $R_1$ and $R_2$ because we need to correlate the row and column indices in the sum. This ismorphism 
amounts to a choice of basis and is specified by requiring for $\sigma$ in the subgroup of which $R_1$ and $R_2$
are irreducible representations, we have $\Gamma_{R_1}(\sigma)=\Gamma_{R_2}(\sigma)$. We represent these subspaces
graphically by drawing $R$ as a Young diagram and placing two labels in each box to be dropped. If a total of
$m$ boxes are to be dropped the labels run from 1 to $m$. To get the row (column) subspace $R_1$ ($R_2$)
drop boxes from $R$ according to the upper (lower) index in each box. 

Looking back at the cutting
rules, it is clear that we only need to compute restricted characters of cycles $(i_1 i_2 \cdots i_k)$ for the case
that the indices $i_1, i_2, \cdots i_k$ are associated to dropped boxes, i.e. they are left inert by the subgroup
whose carrier space we trace over. We have this in mind for the remainder of this appendix. The general algorithm 
used to compute these restricted characters has three steps:

\begin{itemize}

\item{} Decompose the group element whose trace is to be computed into a product of two cycles of the form
        $\Gamma_R\left((i,i+1)\right)$. Insert a complete set of states between each factor.

\item{} The only non-zero matrix elements of each $\Gamma_R\left((i,i+1)\right)$ factor, are obtained when the order
        of boxes dropped to obtain the carrier space of the bra matches the order of boxes dropped to obtain the carrier 
        space of the ket, except for the $(n-i+1)^{\rm th}$ and $(n-i+2)^{\rm th}$ boxes, whose order can be swapped.

\item{} The known value of the matrix elements for precisely the two cases arising in the previous point
        are plugged in to get the value of the restricted character.

\end{itemize} 

A very convenient way to implement this algorithm is by using {\it strand diagrams} \cite{Bekker:2007ea}. 
If, after factorizing the group element as described in the first point above, $n$ indices are involved,
we draw a picture with $n$ columns. The columns are populated by labeled strands - each strand
represents one of the boxes that are to be dropped. Label the strands by the upper index in the box. The box that appears in 
the first column is to be dropped first; the box in the second column is to be dropped second and so on.  The strands
are ordered at the top of the diagram, according to the order in which they must be dropped to get the row index. The strands
are ordered at the bottom of the diagram according to the column index. The strands move from the top of the diagram to the 
bottom of the diagram, without breaking, so that strand ends at the top connect to the corresponding strand ends at the bottom.
To connect the strands (which in general are in a different order at the top and bottom of the diagram) we need to weave the
strands, thereby allowing them to swap columns. The allowed swaps depends on the specific group element whose trace we are
computing. To determine the allowed swaps, write the group element as a product of cycles of the form $(i,i+1)$. Each cycle $(i,i+1)$ 
is drawn as a box which straddles the columns $i$ and $i+1$. Boxes on the right are drawn above boxes on the left. When the strands 
pass through a box, they may do so without swapping or by swapping columns. Each box is associated with a factor. Imagine 
that the strands passing through the box, reading from left to right, are labeled $n$ and $m$. The weights associated with 
these boxes are $c_n$ and $c_m$ respectively. If the strands do not swap inside the box the factor for the box is
$$ f_{\rm no\,\, swap}={1\over c_n-c_m}.$$
If the strands do swap inside the box, the factor is
$$ f_{\rm swap}=\sqrt{1-{1\over (c_n-c_m)^2}}.$$
Denote the product of the factors, one from each box, by $F$. We have
$$ \Tr_{R_1,R_2}\Big(\Gamma_R(\sigma )\Big) = \sum_i F_i{\rm dim}_{R_1}, $$
where the index $i$ runs over all possible paths consistent with the boundary conditions.

With a little thought, the astute reader should be able to convince herself that this graphical rule is nothing
but a convenient representation of the algorithm given above. We end with an example. The character
$$ \chi_1=\Tr_{\young({\,}{\,}{\onethree},{\,}{\twoone},{\threetwo})}\Big(\Gamma_{\yng(3,2,1)}\big((6,4)\big)\Big).$$
is represented by the strand diagram of figure 6.
\myfig{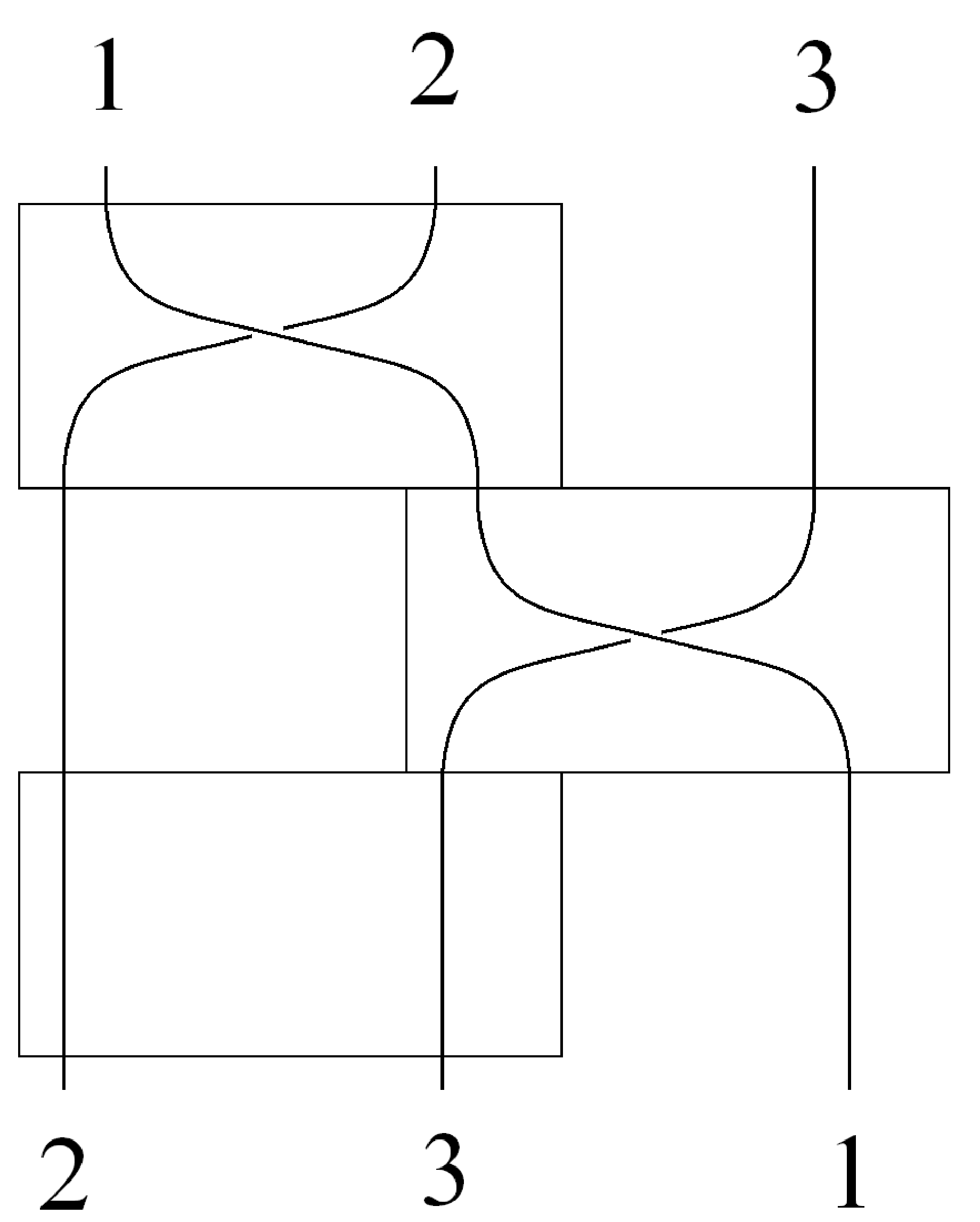}{3.0}{The strand diagram used in the computation of $\chi_1$.}
To obtain this strand diagram write $(6,4)=(6,5)(4,5)(6,5)$. The factors for the upper most, middle
and lower most boxes are $\sqrt{1-{1\over (c_1-c_2)^2}}$, $\sqrt{1-{1\over (c_1-c_3)^2}}$,
and ${1\over c_2-c_3}$ respectively. Thus,
\begin{eqnarray}
\chi_1&=&\sqrt{1-{1\over (c_1-c_2)^2}}\sqrt{1-{1\over (c_1-c_3)^2}}{1\over c_2-c_3}{\rm dim}_{\yng(2,1)}
\nonumber\\
&=&2\sqrt{1-{1\over (c_1-c_2)^2}}\sqrt{1-{1\over (c_1-c_3)^2}}{1\over c_2-c_3}.
\nonumber
\end{eqnarray}
For further details and more examples, see \cite{Bekker:2007ea}.

\section{Example Correlator}

In this appendix we give the details of the computation of a correlator of the type considered in section 2.2
$$ I_{RR_1,RR_1}=\left\langle \chi^{(1)}_{R,R_1}(\chi^{(1)}_{R,R_1})^\dagger\right\rangle\, .$$
We deal with three impurities in the open string $W^i_j=(YZ^3 Y)^i_j$ and take $R_1$ to be the rectangular Young
diagram with $N$ rows and $M$ columns with $M=O(N)$. $R$ is given by adding a box in the upper right hand corner,
i.e. in the first row. This example is already involved enough to nicely illustrate the use of our technology. 

{\vskip 0.5cm}

\noindent
{\it No Brane/String Contractions:} This contribution comes from the diagram given below.
\myfig{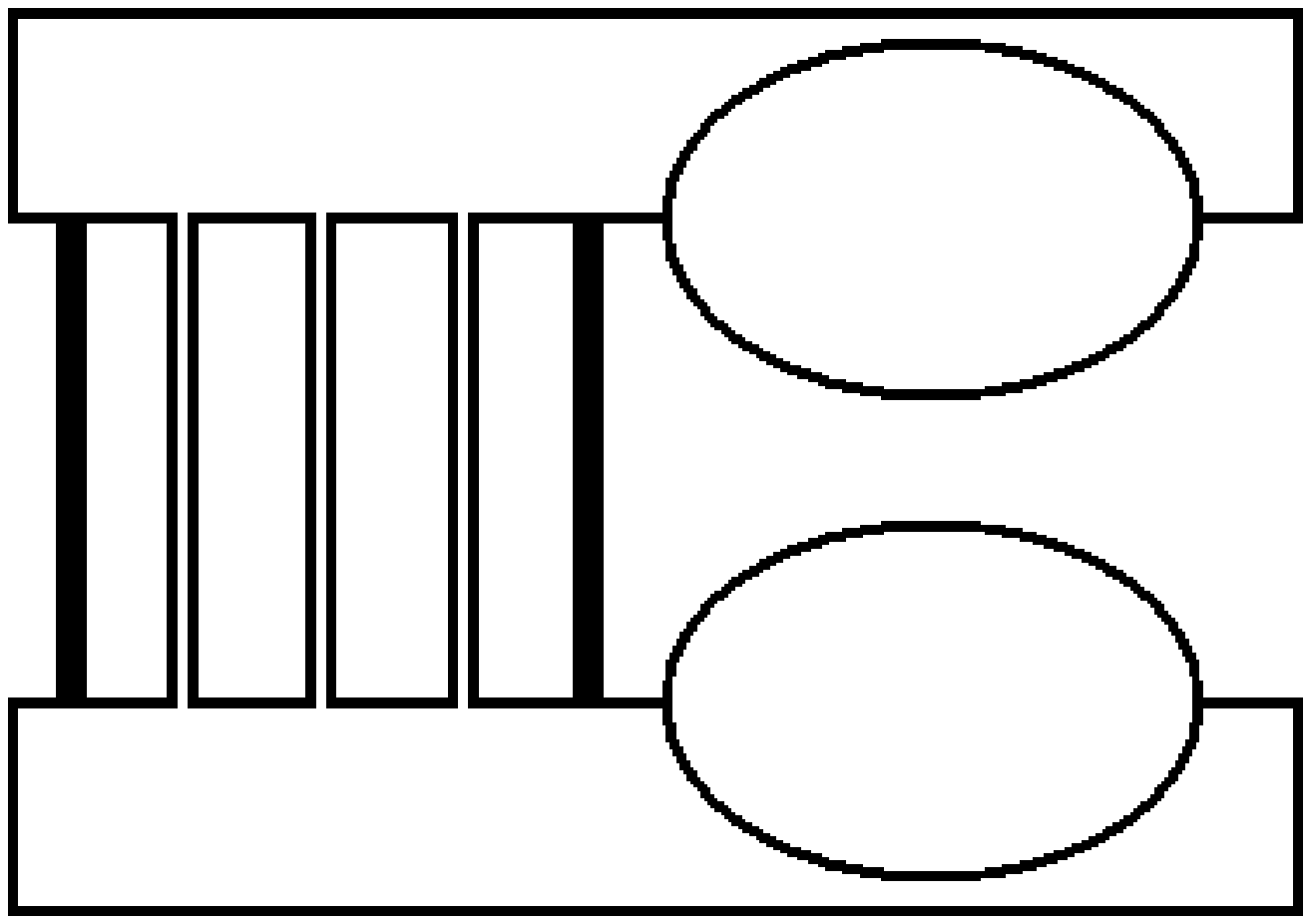}{4.5}{The contribution with no brane/string contractions.}

Using the rules of \cite{deMelloKoch:2007uu} we easily obtain, at leading order in a large $N$ expansion
$$ I_{RR_1,RR_1}^{(0)}=N^4 {{\rm hooks}_R\over {\rm hooks}_{R'}}f_R=N^3 (M+N) f_R \, .$$

{\vskip 0.5cm}

\noindent
{\it One Brane/String Contraction:} This contribution comes from the three diagrams given below.
\myfig{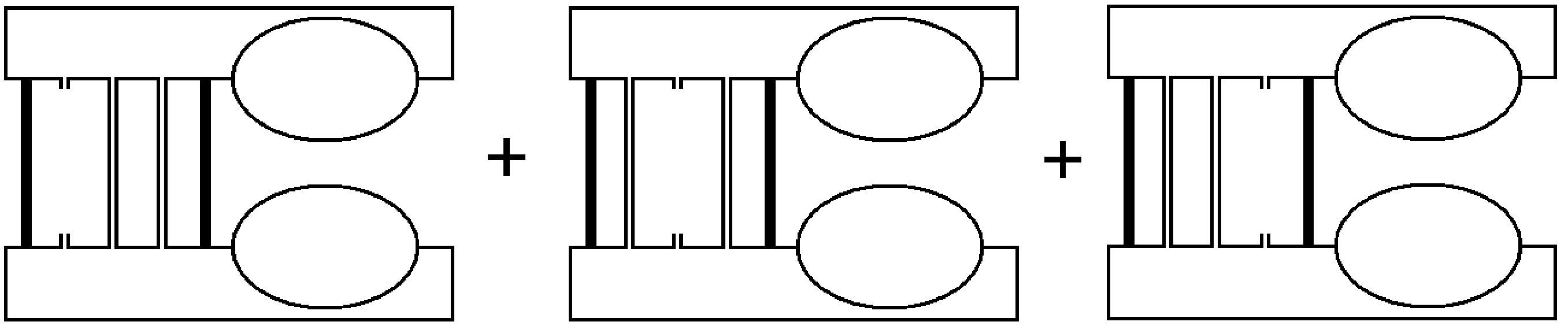}{12.0}{The contribution with one brane/string contraction.}

All three diagrams give the same contibution. We do not need to use our cutting rules yet; we do use the results
of Appendices B.1 and B.2. The result is
$$ I_{RR_1,RR_1}^{(1)}=3N^2 \left\langle 
\Tr \left({d\over dZ}{d\over dZ^\dagger}\right)
\Tr \left({d\over dW}{d\over dW^\dagger}\right)
\chi^{(1)}_{R,R_1}(\chi^{(1)}_{R,R_1})^\dagger\right\rangle$$
$$=3N^2 (MN) {{\rm hooks}_R\over {\rm hooks}_{R'}}f_R=3M N^2 (M+N) f_R \, .$$

{\vskip 0.5cm}

\noindent
{\it Two Brane/String Contractions:} This contribution comes from the three diagrams given below.
\myfig{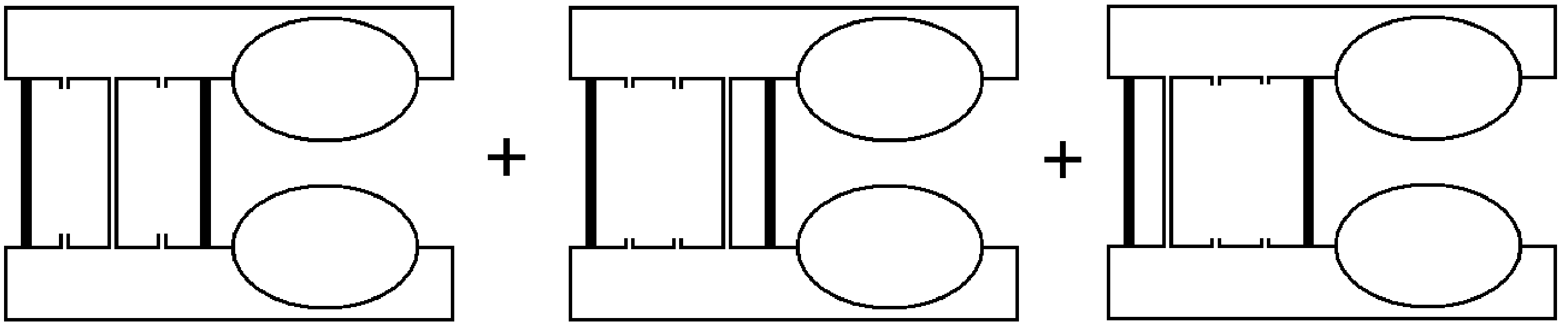}{12.0}{The contribution with two brane/string contractions.}

The first diagram is the simplest to evaluate. We can again do it without using the cutting rules.
The result is
$$ \left\langle 
\Tr \left({d\over dZ}{d\over dZ^\dagger}\right)^2
\Tr \left({d\over dW}{d\over dW^\dagger}\right)
\chi^{(1)}_{R,R_1}(\chi^{(1)}_{R,R_1})^\dagger\right\rangle$$
$$=(MN)^2 {{\rm hooks}_R\over {\rm hooks}_{R'}}f_R= M^2 N (M+N) f_R \, .$$
The evaluation of the second and third diagrams are exactly the same. Consider the second diagram. We need
to evaluate
$$ N \left\langle 
\Tr \left({d\over dZ}{d\over dZ}{d\over dZ^\dagger}{d\over dZ^\dagger}\right)
\Tr \left({d\over dW}{d\over dW^\dagger}\right)
\chi^{(1)}_{R,R_1}(\chi^{(1)}_{R,R_1})^\dagger\right\rangle\, .$$
We now need to use our cutting rules and the associated open string holders. We start to use the graphical notation that draws the 
Young diagram, with the open string word ($w$) and the open string place holders ($1$ and $2$) on $R$ (see Appendix C.1 and G).
We draw $R_1$ with 5 rows and 5 columns, but our results hold for general $M$ and $N$. After cutting we have to evaluate
{\small
$$
\chi_A= \Tr\left({d\over dX_1}\right) \left[
 \tilde{\chi}_{\young({\,}{\,}{\,}{\,}{\,}{w},{\,}{\,}{\,}{\,}{\,},{\,}{\,}{\,}{\,}{\,},{\,}{\,}{\,}{\,}{2},{\,}{\,}{\,}{\,}{1})}
+\tilde{\chi}_{\young({\,}{\,}{\,}{\,}{\,}{w},{\,}{\,}{\,}{\,}{\,},{\,}{\,}{\,}{\,}{\,},{\,}{\,}{\,}{\,}{\,},{\,}{\,}{\,}{2}{1})}
\right].$$}
The tilde on $\chi$ is to denote the fact that there is a trace insertion factor of $(n-1,n-2)$ arising from the cutting.
Using strand diagrams we can eliminate the trace insertion factors for each term
{\small
$$ \tilde{\chi}_{\young({\,}{\,}{\,}{\,}{\,}{w},{\,}{\,}{\,}{\,}{\,},{\,}{\,}{\,}{\,}{\,},{\,}{\,}{\,}{\,}{2},{\,}{\,}{\,}{\,}{1})}
=-\chi_{\young({\,}{\,}{\,}{\,}{\,}{w},{\,}{\,}{\,}{\,}{\,},{\,}{\,}{\,}{\,}{\,},{\,}{\,}{\,}{\,}{2},{\,}{\,}{\,}{\,}{1})},\qquad
\tilde{\chi}_{\young({\,}{\,}{\,}{\,}{\,}{w},{\,}{\,}{\,}{\,}{\,},{\,}{\,}{\,}{\,}{\,},{\,}{\,}{\,}{\,}{\,},{\,}{\,}{\,}{2}{1})}
=\chi_{\young({\,}{\,}{\,}{\,}{\,}{w},{\,}{\,}{\,}{\,}{\,},{\,}{\,}{\,}{\,}{\,},{\,}{\,}{\,}{\,}{\,},{\,}{\,}{\,}{2}{1})} \, .$$}
After using the subgroup swap rule to swap $w$ and $X_1$, we can compute the reduction to obtain (there are some terms that arise from
swapping $w$ and 1; these are however $O\left({1\over N^2}\right)$ so they can be dropped to leading order in $N$)
{\small
$$
\chi_A= -M\chi_{\young({\,}{\,}{\,}{\,}{\,}{w},{\,}{\,}{\,}{\,}{\,},{\,}{\,}{\,}{\,}{\,},{\,}{\,}{\,}{\,}{2},{\,}{\,}{\,}{\,})}
+M\chi_{\young({\,}{\,}{\,}{\,}{\,}{w},{\,}{\,}{\,}{\,}{\,},{\,}{\,}{\,}{\,}{\,},{\,}{\,}{\,}{\,}{\,},{\,}{\,}{\,}{2})}$$
To get the contribution from the second diagram, we now simply need to compute 
$$N\left\langle \Tr \left({d\over dX_2}{d\over dX_2^\dagger}\right)
\Tr \left({d\over dW}{d\over dW^\dagger}\right)
\chi_A\chi^\dagger_A \right\rangle .$$ 
To obtain this, we need to use
$$ 
h_1={{\rm hooks}_{\tiny \young({\,}{\,}{\,}{\,}{\,}{\,},{\,}{\,}{\,}{\,}{\,},{\,}{\,}{\,}{\,}{\,},{\,}{\,}{\,}{\,}{\,},{\,}{\,}{\,}{\,})}
\over
{\rm hooks}_{\tiny \young({\,}{\,}{\,}{\,}{\,},{\,}{\,}{\,}{\,}{\,},{\,}{\,}{\,}{\,}{\,},{\,}{\,}{\,}{\,},{\,}{\,}{\,}{\,})}
}={M(M+N)\over 2},\qquad
h_2={{\rm hooks}_{\tiny \young({\,}{\,}{\,}{\,}{\,}{\,},{\,}{\,}{\,}{\,}{\,},{\,}{\,}{\,}{\,}{\,},{\,}{\,}{\,}{\,}{\,},{\,}{\,}{\,}{\,})}
\over
{\rm hooks}_{\tiny \young({\,}{\,}{\,}{\,}{\,},{\,}{\,}{\,}{\,}{\,},{\,}{\,}{\,}{\,}{\,},{\,}{\,}{\,}{\,}{\,},{\,}{\,}{\,})}
}={M(M+N)\over 2}.$$
It is now straight forward to obtain
$$N\left\langle \Tr \left({d\over dX_2}{d\over dX_2^\dagger}\right)
\Tr \left({d\over dW}{d\over dW^\dagger}\right)
\chi_A\chi^\dagger_A \right\rangle =N M f_R\left(h_1+h_2\right)$$
$$=N M^2 (M+N) f_R\, .$$ 
Notice that although the computation for diagram 2 was completely different to the computation for diagram 1, they give exactly the same
result. As already mentioned, the third diagram gives exactly the same contribution as the second so that
$$ I_{RR_1,RR_1}^{(2)}=3M^2 N (M+N) f_R \, .$$

{\vskip 0.25cm}

\noindent
{\it Three Brane/String Contractions:} This contribution comes from the diagram given below.
\myfig{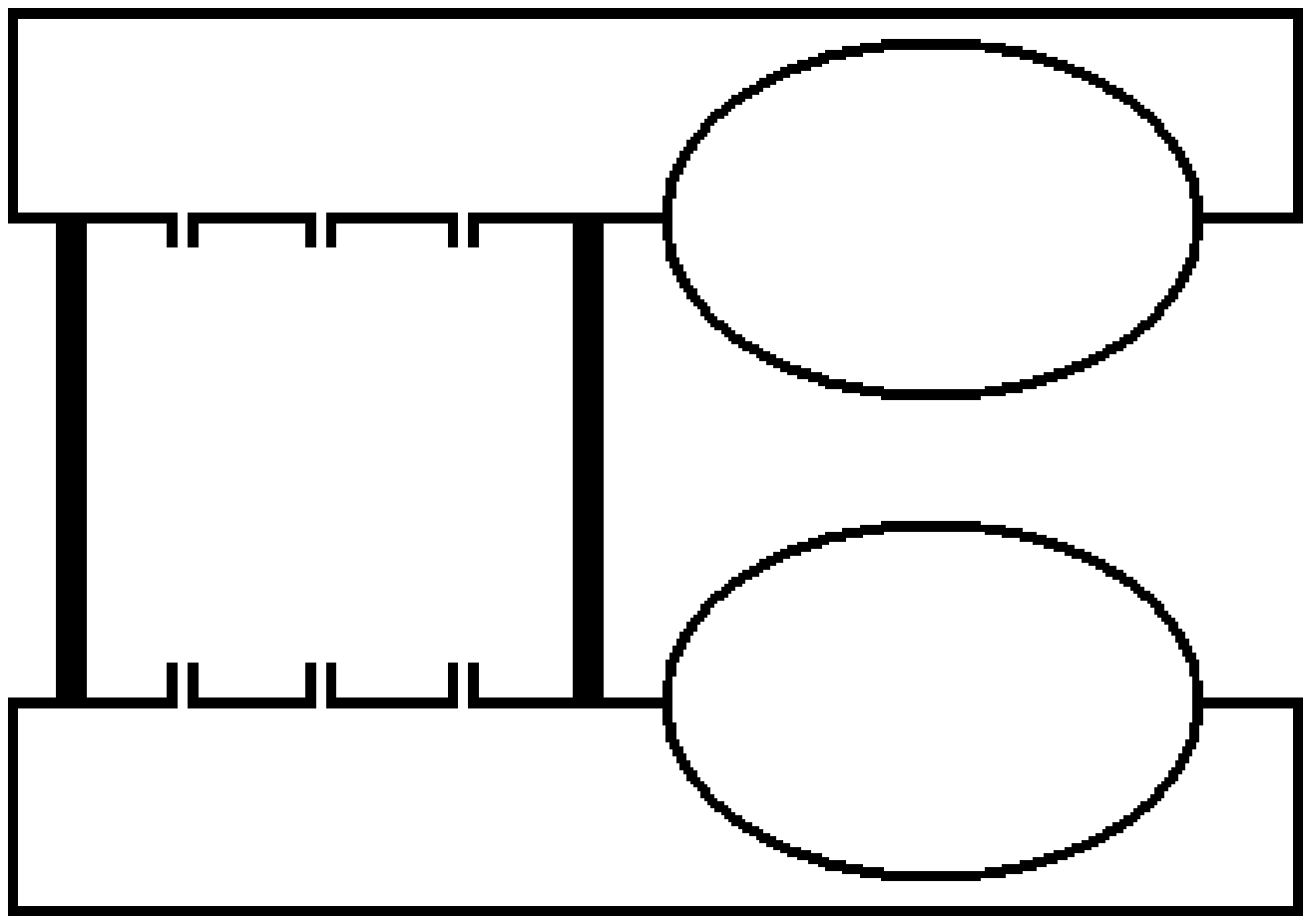}{4.0}{The contribution with three brane/string contractions.}

For this contribution we need to evaluate
$$ \left\langle 
\Tr \left({d\over dZ}{d\over dZ}{d\over dZ}{d\over dZ^\dagger}{d\over dZ^\dagger}{d\over dZ^\dagger}\right)
\Tr \left({d\over dW}{d\over dW^\dagger}\right)
\chi^{(1)}_{R,R_1}(\chi^{(1)}_{R,R_1})^\dagger\right\rangle\, .$$
We cut two holomorphic derivatives and two antiholomorphic derivatives out of the trace. Thus, we will need a total of
three open string place holders; the trace insertion factor is $(n-3,n-2)(n-1,n-2)$. To recover the trace over $R_1$ we
again need to sum over all ways of distributing the open string place holders. The result is
{\small
$$
 \tilde{\chi}_{\young({\,}{\,}{\,}{\,}{\,}{w},{\,}{\,}{\,}{\,}{\,},{\,}{\,}{\,}{\,}{\,},{\,}{\,}{\,}{\,}{3},{\,}{\,}{\,}{2}{1})}
+\tilde{\chi}_{\young({\,}{\,}{\,}{\,}{\,}{w},{\,}{\,}{\,}{\,}{\,},{\,}{\,}{\,}{\,}{\,},{\,}{\,}{\,}{\,}{\,},{\,}{\,}{3}{2}{1})}
+\tilde{\chi}_{\young({\,}{\,}{\,}{\,}{\,}{w},{\,}{\,}{\,}{\,}{\,},{\,}{\,}{\,}{\,}{3},{\,}{\,}{\,}{\,}{2},{\,}{\,}{\,}{\,}{1})}
+\tilde{\chi}_{\young({\,}{\,}{\,}{\,}{\,}{w},{\,}{\,}{\,}{\,}{\,},{\,}{\,}{\,}{\,}{\,},{\,}{\,}{\,}{\,}{2},{\,}{\,}{\,}{3}{1})} \, .$$}
After accounting for the trace insertion factor, we obtain
{\small
$$
-{1\over 2}\chi_{\young({\,}{\,}{\,}{\,}{\,}{w},{\,}{\,}{\,}{\,}{\,},{\,}{\,}{\,}{\,}{\,},{\,}{\,}{\,}{\,}{3},{\,}{\,}{\,}{2}{1})}
+{\sqrt{3}\over 2}\chi_{\young({\,}{\,}{\,}{\,}{\,}{w},{\,}{\,}{\,}{\,}{\,},{\,}{\,}{\,}{\,}{\,},{\,}{\,}{\,}{\,}{\threetwo},{\,}{\,}{\,}{\twothree}{1})}
+\chi_{\young({\,}{\,}{\,}{\,}{\,}{w},{\,}{\,}{\,}{\,}{\,},{\,}{\,}{\,}{\,}{\,},{\,}{\,}{\,}{\,}{\,},{\,}{\,}{3}{2}{1})}$$
$$+\chi_{\young({\,}{\,}{\,}{\,}{\,}{w},{\,}{\,}{\,}{\,}{\,},{\,}{\,}{\,}{\,}{3},{\,}{\,}{\,}{\,}{2},{\,}{\,}{\,}{\,}{1})}
-{1\over 2}\chi_{\young({\,}{\,}{\,}{\,}{\,}{w},{\,}{\,}{\,}{\,}{\,},{\,}{\,}{\,}{\,}{\,},{\,}{\,}{\,}{\,}{2},{\,}{\,}{\,}{3}{1})}
-{\sqrt{3}\over 2}\chi_{\young({\,}{\,}{\,}{\,}{\,}{w},{\,}{\,}{\,}{\,}{\,},{\,}{\,}{\,}{\,}{\,},{\,}{\,}{\,}{\,}{\twothree},{\,}{\,}{\,}{\threetwo}{1})} \, .$$}
We now need to use the subgroup swap rule so that we can reduce with respect to $X_1$ and $X_2$. There is again a dramatic simplification
because the terms in which the location of $w$ changes are suppressed at large $N$. The result after reducing is
{\small
$$
-{M^2\over 2}\chi_{\young({\,}{\,}{\,}{\,}{\,}{w},{\,}{\,}{\,}{\,}{\,},{\,}{\,}{\,}{\,}{\,},{\,}{\,}{\,}{\,}{3},{\,}{\,}{\,})}
+M^2\chi_{\young({\,}{\,}{\,}{\,}{\,}{w},{\,}{\,}{\,}{\,}{\,},{\,}{\,}{\,}{\,}{\,},{\,}{\,}{\,}{\,}{\,},{\,}{\,}{3})}
+M^2 \chi_{\young({\,}{\,}{\,}{\,}{\,}{w},{\,}{\,}{\,}{\,}{\,},{\,}{\,}{\,}{\,}{3},{\,}{\,}{\,}{\,},{\,}{\,}{\,}{\,})}
-{M^2\over 2}\chi_{\young({\,}{\,}{\,}{\,}{\,}{w},{\,}{\,}{\,}{\,}{\,},{\,}{\,}{\,}{\,}{\,},{\,}{\,}{\,}{\,},{\,}{\,}{\,}{3})} \, .$$}
To get the contribution from the three brane/string contractions, we now need to compute
$$ \left\langle\chi_A\chi_A^\dagger\right\rangle = M^3 (M+N) f_R .$$
To get this we used
$$ 
{{\rm hooks}_{\tiny \young({\,}{\,}{\,}{\,}{\,}{\,},{\,}{\,}{\,}{\,}{\,},{\,}{\,}{\,}{\,}{\,},{\,}{\,}{\,}{\,}{\,},{\,}{\,}{\,})}
\over
{\rm hooks}_{\tiny \young({\,}{\,}{\,}{\,}{\,},{\,}{\,}{\,}{\,}{\,},{\,}{\,}{\,}{\,}{\,},{\,}{\,}{\,}{\,},{\,}{\,}{\,})}}
={2M(M+N)\over 3},\qquad
{{\rm hooks}_{\tiny \young({\,}{\,}{\,}{\,}{\,}{\,},{\,}{\,}{\,}{\,}{\,},{\,}{\,}{\,}{\,}{\,},{\,}{\,}{\,}{\,}{\,},{\,}{\,}{\,})}
\over
{\rm hooks}_{\tiny \young({\,}{\,}{\,}{\,}{\,},{\,}{\,}{\,}{\,}{\,},{\,}{\,}{\,}{\,}{\,},{\,}{\,}{\,}{\,}{\,},{\,}{\,})}}
={M(M+N)\over 3},$$
$$ 
{{\rm hooks}_{\tiny \young({\,}{\,}{\,}{\,}{\,}{\,},{\,}{\,}{\,}{\,}{\,},{\,}{\,}{\,}{\,}{\,},{\,}{\,}{\,}{\,},{\,}{\,}{\,}{\,})}
\over
{\rm hooks}_{\tiny \young({\,}{\,}{\,}{\,}{\,},{\,}{\,}{\,}{\,}{\,},{\,}{\,}{\,}{\,}{\,},{\,}{\,}{\,}{\,},{\,}{\,}{\,})}}
={2M(M+N)\over 3},\qquad
{{\rm hooks}_{\tiny \young({\,}{\,}{\,}{\,}{\,}{\,},{\,}{\,}{\,}{\,}{\,},{\,}{\,}{\,}{\,}{\,},{\,}{\,}{\,}{\,},{\,}{\,}{\,}{\,})}
\over
{\rm hooks}_{\tiny \young({\,}{\,}{\,}{\,}{\,},{\,}{\,}{\,}{\,}{\,},{\,}{\,}{\,}{\,},{\,}{\,}{\,}{\,},{\,}{\,}{\,}{\,})}}
={M(M+N)\over 3}.$$

Putting things together, we have
$$I_{RR_1,RR_1}=(N^3 + 3M N^2 + 3M^2 N + M^3) (M+N)f_R = \left(1+{M\over N}\right)^3 N^3 (M+N)f_R\, .$$

\section{Exact Results for the Annulus}

In this appendix we consider a background $\chi_B(Z)$ where $B$ is a Young diagram with $M$ columns and $N$ rows. We will compute
the two correlators
$$ I_1={\langle\chi_B\chi_B^\dagger \Tr (Z^nZ^{\dagger\, n}) \rangle \over \langle\chi_B\chi_B^\dagger \rangle},$$
and
$$ I_2={\langle\Tr \left( {d^n\over dZ^n}{d^n\over dZ^{\dagger\, n}}\right)\chi_B\chi_B^\dagger \rangle \over \langle\chi_B\chi_B^\dagger \rangle},$$
in the large $N$ limit.

\subsection{Computation of $I_1$}

We will make use of a dummy field $D$ which does not interact with $Z$ and has a two point function
$$\left\langle (D^\dagger)^k_l D^i_j\right\rangle = \delta^i_l\delta^k_j ,$$ 
Including $D$ does not change the value of any normalized correlation functions of operators
built only out of $Z$ and $Z^\dagger$. In particular, it does not change the value of $I_1$.
Using $D$, we can rewrite
$$ I_1={\langle\chi_B\chi_B^\dagger \Tr (Z^n D)\Tr (D^\dagger Z^{\dagger\, n}) \rangle 
                      \over \langle\chi_B\chi_B^\dagger \rangle}.$$
This is a useful step, because after using the identities
$$ \Tr (Z^n D)={1\over n+1}D_{ij}{d\over dZ_{ij}}\Tr (Z^{n+1}),$$
and (this identity was proved in Appendix 6 of \cite{Corley:2002mj})
$$ \Tr (Z^{n+1})=\sum_{s=0}^n (-1)^s\chi_{(n+1-s,1^s)}(Z),$$
where $(n+1-s,1^s)$ denotes a Young diagram with $s+1$ rows; the first row has $n+1-s$ boxes and all
remaining rows have one box, we can write $\Tr (Z^n D)$ as a sum of restricted Schur polynomials
$$ \Tr (Z^n D) = {1\over n+1}\sum_{s=0}^n\sum_{h_s}(-1)^s\chi_{(n+1-s,1^s),h_s}(Z,D),$$
where $h_s$ is an irreducible representation of $S_n$. The sum over $h_s$ is a sum over all possible $S_n$ 
irreducible representations that can be suduced from the $S_{n+1}$ representation $(n+1-s,1^s)$. To proceed,
we would like to evaluate the product
\begin{equation}
\chi_B(Z)\chi_{(n+1-s,1^s),h_s}(Z,D)
\label{prod}
\end{equation}
for any $s$ and $h_s$. This product can be computed using the restricted Littlewood-Richardson rule derived
in the second of \cite{Bhattacharyya:2008rb}. The difficult part of this computation entails evaluating the
restricted Littlewood-Richardson numbers, which include the sum
$$\sum_{\sigma_1\in S_{NM}}\sum_{\sigma_2\in S_{n+1}}
\chi_B(\sigma_1 )\chi_{(n+1-s,1^s),h_s}(\sigma_2 )\chi_{R,R'}(\sigma_1\circ\sigma_2).$$
To evaluate this sum, note that both
$$ {d_B\over NM!}\sum_{\sigma_1\in S_{NM}} \chi_B(\sigma_1 )\sigma_1\qquad {\rm and}\qquad
{d_{(n+1-s,1^s)}\over (n+1)!}\sum_{\sigma_2\in S_{n+1}}\chi_{(n+1-s,1^s),h_s}(\sigma_2 ) \sigma_2 $$
are projection operators. Thus, the sum we need to compute is simply the partial trace (over $(R,R')$) of the
direct product of two projectors. In general this is not a very useful observation because one can't choose
a basis which is both simultaneously a basis of $B$ and $((n+1-s,1^s),h_s)$ on the one hand and $(R,R')$ on
the other. However, for the case we consider here a simultaneous basis can indeed be chosen as we now
explain.

The above sum is needed to compute the coefficient of the term $\chi_{R,R'}(Z,D)$ appearing in
the product (\ref{prod}). Since $B$ has $N$ rows, we can only stack $((n+1-s,1^s),h_s)$ 
as a complete Young diagram, to the right of $B$; denote this new Young diagram by $(+(n+1-s,1^s),+h_s)$.
To see that this is the case, note that we could start with $(R,R')$ which is an irreducible representation
of $S_{NM+n+1}$ and keep restricting to smaller and smaller subgroups, by freezing the indices that $S_{n+1}$
acts on. Doing $n+1$ restrictions we have the subgroup $S_{NM}$ and we must have reduced $(R,R')$ to $B$.
This forces $(R,R')$ to be $(+(n+1-s,1^s),+h_s)$ and it provides a simultaneous basis for
$B$ and $((n+1-s,1^s),h_s)$ and for $(+(n+1-s,1^s),+h_s)$. 

It is now straight forward to see that
$$
{d_B\over NM!}{d_{(n+1-s,1^s)}\over (n+1)!}\sum_{\sigma_1\in S_{NM}}\sum_{\sigma_2\in S_{n+1}}
\chi_B(\sigma_1 )\chi_{(n+1-s,1^s),h_s}(\sigma_2 )\chi_{+(n+1-s,1^s),+h_s}(\sigma_1\circ\sigma_2)
=d_{h_s}d_B,$$
where the right hand side is nothing but the dimension of the space that we traced over. Consequently,
$$\sum_{\sigma_1\in S_{NM}}\sum_{\sigma_2\in S_{n+1}}
\chi_B(\sigma_1 )\chi_{(n+1-s,1^s),h_s}(\sigma_2 )\chi_{R,R'}(\sigma_1\circ\sigma_2)
=(n+1)!NM! {d_{h_s}\over d_{(n+1-s,1^s)} }.$$
Some straightforward manipulations now give
$$\chi_B(Z)\Tr (Z^n D)={1\over n+1}{N\over N+M}
\sum_{s=0}^n\sum_{h_s}(-1)^s\chi_{+(n+1-s,1^s),+h_s}(Z,D). $$
Thus, we have reduced the computation of $I_1$ to the computation of a two point function which is easily
performed (we keep only the leading term at large $N$)
\begin{eqnarray}
\nonumber
I_1 &=& 
{\langle\chi_B\chi_B^\dagger \Tr (Z^nZ^{\dagger\, n}) \rangle \over \langle\chi_B\chi_B^\dagger \rangle}\cr
\nonumber
&=&{1\over f_B}{1\over (n+1)^2}{N^2\over (N+M)^2}
\langle\sum_{s=0}^n\sum_{h_s}(-1)^s\chi_{+(n+1-s,1^s),+h_s}(Z,D)
\sum_{t=0}^n\sum_{h_t}(-1)^t\chi_{+(n+1-t,1^t),+h_s}(Z,D)^\dagger\rangle\cr
\nonumber
&=&{1\over (n+1)^2}N(M+N)^n\sum_{s=0}^n\sum_{h_s} {({\rm hooks})_{(n+1-s,1^s)}\over ({\rm hooks})_{h_s}}\cr
\nonumber
&=& N(M+N)^n\, .
\end{eqnarray}

\subsection{Computation of $I_2$}

It is clear that we can write
$$ I_2={\langle\chi_B\chi_B^\dagger :\Tr (Z^nZ^{\dagger\, n}): \rangle \over \langle\chi_B\chi_B^\dagger \rangle}
\equiv\langle :\Tr (Z^nZ^{\dagger\, n}): \rangle_B ,$$
where $:O:$ denotes the normal ordering of $O$. Thus, we can obtain $I_2$ from $I_1$ by subtracting
all terms with an odd number of self contractions (contractions between two fields in 
$\Tr (Z^nZ^{\dagger\, n})$) from $I_1$ and adding
back all the terms with an even number of self contractions. The term with
one self contraction, for example, gives
$$\sum_{r=1}^n \langle\Tr (Z^{n-r}(Z^\dagger)^{n-r}) \Tr (Z^{r-1}(Z^\dagger)^{r-1})\rangle_B
=\sum_{r=1}^n \langle\Tr (Z^{n-r}(Z^\dagger)^{n-r})\rangle_B \langle\Tr (Z^{r-1}(Z^\dagger)^{r-1})\rangle_B$$
$$ = \sum_{r=1}^n N(M+N)^{n-r}N(M+N)^{r-1}=n N^2 (M+N)^{n-1}.$$
To obtain this result we made use of large $N$ factorization and the result of the previous subsection. 
A very similar argument gives
$$ {n!\over c!(n-c)!}N^{1+c}(N+M)^{n-c} $$
for the term with $c$ self contractions. Thus
\begin{eqnarray}
\nonumber 
I_2 &=& N(N+M)^n-\sum_{c=1}^{n} {n!\over c!(n-c)!}(-N)^{1+c}(N+M)^{n-c}\cr
\nonumber 
&=& N \sum_{c=0}^{n} {n!\over c!(n-c)!}(-N)^{c}(N+M)^{n-c}\cr
\nonumber 
&=& N(N+M-N)^n\cr
&=& NM^n \, .
\end{eqnarray}

\section{Last Site Dictionary}

In this section we will explain how to translate between a ``closed string'' description of the operator
$$ w=\Tr (YZ^{n_1}YZ^{n_2}Y\cdots YZ^{n_L}) $$
and an ``open string'' description
$$ \sum_{R,R'}\alpha_{R,R'}\chi_{R,R'}^{(1)}(Z,w)\qquad w^i_j=(YZ^{n_1}YZ^{n_2}Y\cdots YZ^{n_{L-1}}Y)^i_j ,$$
where in this second description the last site is described by the Young diagrams $R,R'$. One simply makes repeated
use of the identity
$$ \chi^{(1)}_{R,R'}(Z,w)-\chi_{R'}(Z)\Tr (w) =
\sum_\alpha {1\over d_{R_\alpha''}}\Tr_{R''_\alpha}(\Gamma_R\left[ (n,n-1)\right])
\chi^{(1)}_{R',R''_\alpha}(Z,Zw) .$$
which was derived in \cite{deMelloKoch:2007uv}. The second term on the LHS in the above identity does not contribute at large $N$.
Start from
$$ \chi^{(1)}_{{\tiny\yng(1)},\cdot}(Z,Z^{n_L}w)\equiv \Tr (Z^{n_L}w),$$
and use the identity to pull $Z$'s off $Z^{n_L}$ and onto the Young diagram $R$. For example, for $n_L=1,2,3$ we have
$$ \Tr (Zw)={1\over 2}\left(\chi_{\tiny\young({\,}{w})}-\chi_{\tiny\young({\,},{w})}\right),$$
$$ \Tr (Z^2 w)={1\over 3}\left(\chi_{\tiny\young({\,}{\,}{w})}-\chi_{\tiny\young({\,}{\,},{w})}
-\chi_{\tiny\young({\,}{w},{\,})}+\chi_{\tiny\young({\,},{\,},{w})}\right),$$
$$ \Tr (Z^3 w)={1\over 4}\left(
\chi_{\tiny\young({\,}{\,}{\,}{w})}
-\chi_{\tiny\young({\,}{\,}{\,},{w})}-\chi_{\tiny\young({\,}{\,}{w},{\,})}
+\chi_{\tiny\young({\,}{\,},{\,},{w})}+\chi_{\tiny\young({\,}{w},{\,},{\,})}
-\chi_{\tiny\young({\,},{\,},{\,},{w})}\right).$$
These formulas are exact.

\section{Notation}

In this appendix, we review the definition of the restricted Schur polynomial; for more details consult
\cite{Balasubramanian:2004nb,deMelloKoch:2007uu,deMelloKoch:2007uv,Bekker:2007ea}. The dual of a giant graviton is a Schur polynomial,
which is labeled by a Young diagram. 
Operators dual to excitations of giant gravitons are obtained by inserting words $(W^{(a)})^j_i$ describing the open strings
(one word for each open string) into the operator describing the system of giant gravitons
\begin{equation}
\chi_{R,R_1}^{(k)}(Z,W^{(1)},...,W^{(k)})={1\over (n-k)!}
\sum_{\sigma\in S_n}\Tr_{R_1}(\Gamma_R(\sigma))\Tr(\sigma Z^{\otimes n-k}W^{(1)}\cdots W^{(k)}),
\label{restrictedschur}
\end{equation}
$$
={1\over (n-k)!}\sum_{\sigma\in S_n}\Tr (\Pi \Gamma_R(\sigma))\Tr(\sigma Z^{\otimes n-k}W^{(1)}\cdots W^{(k)}),
$$
$$\Tr (\sigma Z^{\otimes n-k}W^{(1)}\cdots W^{(k)})= Z^{i_1}_{i_{\sigma (1)}}Z^{i_2}_{i_{\sigma (2)}}\cdots
Z^{i_{n-k}}_{i_{\sigma (n-k)}}(W^{(1)})^{i_{n-k+1}}_{i_{\sigma (n-k+1)}}\cdots
(W^{(k)})^{i_{n}}_{i_{\sigma (n)}}.$$
$\Pi$ is a product of projection operators and/or intertwiners, used to implement the restricted trace. $\Pi$ is defined by
the sequence of irreducible representations used to subduce $R_1$ from $R$ and the chain of subgroups
to which these representations belong. Since the row and column indices of the block that we trace over
(denoted by $R_1$ in the above formula) need not coincide, we need to specify this data separately for both indices. 
Denote the chain of 
subgroups involved in the reduction by ${\cal G}_k\subset {\cal G}_{k-1}\subset\cdots\subset {\cal G}_2\subset {\cal G}_1\subset S_n$.
${\cal G}_m$ is obtained by taking all elements $S_n$ that leave the indices of the strings $W^{(i)}$ with $i\le m$ inert.
To specify the sequence of irreducible representations employed in subducing $R_1$, place a pair of labels into each box,
a lower label and an upper label.
The representations needed to subduce the row label of $R_1$ are obtained by starting with $R$. The second representation
is obtained by dropping the box with upper label equal to 1; the third representation is obtained from the
second by dropping the box with upper label equal to 2 and so on until the box with label $k$ is dropped. The 
representations needed to subduce the column label are obtained in exactly the same way except that instead of using the
upper label, we now use the lower label. For further details see \cite{deMelloKoch:2007uu}.

\end{document}